\documentclass[twocolumn,pra, aps,superscriptaddress,showpacs]{revtex4}

\usepackage{mathptmx}
\usepackage{latexsym}
\usepackage{epstopdf}
\usepackage{color}

\usepackage{graphicx} 
\usepackage{amsmath} 
\usepackage{amssymb} 
\usepackage{amsfonts}
\usepackage{appendix}

\usepackage{ulem} 

\definecolor{mygrey}{gray}{0.35}
\definecolor{myblue}{rgb}{0.,0.,1}
\definecolor{myzard}{cmyk}{0,0,0.05,0}
\definecolor{mywhite}{rgb}{1,1,1}
\definecolor{myred}{rgb}{1,0.,0.3}

\usepackage[colorlinks=true,citecolor=blue,linkcolor=blue]{hyperref}

\def\be{\begin{equation}}
\def\ee{\end{equation}}
\def\ba{\begin{align}}
\def\enda{\end{align}}
\def\bi{\begin{itemize}}
\def\ei{\end{itemize}}

 \def\ee{\mathord{\rm e}}
 
 \def\ii{\mathord{\rm i}}

\def\half{\textstyle\frac{1}{2}}

 \def\ee{\mathord{\rm e}}
 
 \def\ii{\mathord{\rm i}}

\def\half{\textstyle\frac{1}{2}}

\renewcommand{\ii}{{\rm i}}
\renewcommand{\ee}{{\rm e}}

\def\beq{\begin{equation}}
\def\beq{\begin{equation}}
\def\eeq{\end{equation}}

 \newcommand{\ket}[1]{|#1\rangle}
 \newcommand{\bra}[1]{\langle #1|}

\begin{document}


\title{ Dirac-Weyl fermions with arbitrary spin in two-dimensional optical superlattices}

\author{Z. Lan
}
\affiliation{SUPA, Department of Physics, Heriot-Watt University, EH14 4AS, Edinburgh, United Kingdom
}

\author{
N. Goldman
}
\affiliation{Center for Nonlinear Phenomena and Complex Systems - Universit\'e Libre de Bruxelles , 231, Campus Plaine, B-1050 Brussels, Belgium}

\author{A. Bermudez}

\affiliation{
Departamento de F\'isica Te\'orica I,
Universidad Complutense, 
28040 Madrid, 
Spain
}

\affiliation{
Institut f\"ur Theoretische Physik, Albert-Einstein Allee 11, Universit\"at Ulm, 89069 Ulm, Germany
}

\author{W. Lu}
\affiliation{SUPA, Department of Physics, Heriot-Watt University, EH14 4AS, Edinburgh, United Kingdom
}

\author{P. \"Ohberg
}
\affiliation{SUPA, Department of Physics, Heriot-Watt University, EH14 4AS, Edinburgh, United Kingdom
}

\begin{abstract}

Dirac-Weyl fermions are massless relativistic particles with a well-defined helicity which arise in the context of high-energy physics. Here we propose a quantum simulation of these paradigmatic fermions using multicomponent ultracold atoms in a two-dimensional square optical lattice. We find that laser-assisted spin-dependent hopping, specifically tuned to the $(2s+1)$-dimensional representations of the $\mathfrak{su}$(2)
Lie algebra, directly leads to a regime where  the emerging massless excitations correspond to Dirac-Weyl fermions with arbitrary pseudospin $s$. We show that this platform hosts two different phases: a semimetallic phase that occurs for half-integer $s$, and a metallic phase that contains a flat zero-energy band at integer $s$. These phases host a variety of interesting effects, such as a very rich anomalous quantum Hall effect and a remarkable multirefringent Klein tunneling. In addition we show that these effects are  directly related to the number of underlying Dirac-Weyl species and zero modes.  \end{abstract}
\maketitle


\section{Introduction}
 Graphene and topological insulators have recently stimulated an enormous interest   
 at both the theoretical and experimental levels (see~\cite{graphene_rev,top_insulators_rev} and references therein). These two systems share an attractive and unique property, namely, the fact that the electronic transport at low energies is not governed by the usual Schr\"odinger  equation, but rather by its relativistic counterpart: the Dirac equation~\cite{greiner_book}. In graphene,  a single layer of carbon atoms densely packed in a 
honeycomb lattice~\cite{graphene_rev},  the band structure corresponds to a semimetallic phase whose Fermi surface consists of an even number of isolated points. Interestingly, the low-energy excitations around these points display a relativistic dispersion relation, and can be thus described by the two-dimensional Dirac Hamiltonian for massless fermions. Conversely, in three-dimensional topological insulators, which are semiconducting alloys with a strong spin-orbit coupling~\cite{top_insulators_rev},  the bulk band structure corresponds to a gapped insulating phase. Nonetheless, these materials also support a robust surface conductivity which can be described by an odd number of two-dimensional massless Dirac fermions. The recent interest in both graphene and topological insulators is two-fold. On the one hand, they provide concrete platforms where  interesting phenomena were originally predicted in a high-energy context, for example, the Klein paradox~\cite{klein_tunneling}, or axion electrodynamics~\cite{dyn_axion}. On the other hand, they also foresee novel and useful device applications (see for example Ref.~\cite{graphene_device}). The emergence of massless Dirac fermions in solid-state materials  is not only an exciting area of condensed matter, but is also becoming an exciting topic in  cold-atom physics (see e.g.~\cite{dirac_OL}).

In high-energy physics, a relativistic electron is described by a spin-$\half$ fermion, an unavoidable fact that fixes the dimensionality of the Dirac spinor~\cite{greiner_book}. In clear contrast, the pseudospin of an emergent relativistic fermion in solid-state materials depends on the geometry of the underlying lattice. For graphene, the existence of two interpenetrating triangular lattices fixes the dimensionality of the Dirac spinor, which in this case corresponds to a pseudospin-$\half$ fermion. Recently, some researchers have studied alternative lattices, e.g., the $\mathcal{T}_3$ lattice~\cite{weyl_1}, the line-centered-square (Lieb) lattice~\cite{shen:2010, apaja:2010,goldmanlieb},  and the Kagome lattice~\cite{Green2010}, where emergent massless relativistic fermions present a pseudospin-$1$ structure.
These higher-spin relativistic fermions show some
distinctive features with respect to graphene's Dirac fermions, such as all-angle perfect tunneling~\cite{shen:2010}, particle localization \cite{apaja:2010}, and the absence of the anomalous quantum Hall effect \cite{goldmanlieb}. In view of these results, a natural 
question arises: could we engineer an experimental setup where massless relativistic particles with arbitrary spin emerge? Moreover, would these higher spins host novel effects that have no counterpart in the low-spin cases? 

As we try to increase the pseudospin by modifying the lattice geometry, the corresponding structures become increasingly complex. Due to the 
limited number of two-dimensional Bravais lattices and materials that realize such lattices, this construction seems doomed to a failure.  
In this article, we propose an alternative approach which is based on the fact that these pseudospin structures, which arise from complex lattice geometries, can be reproduced by a standard square lattice with a matrix hopping \cite{dirac_fermions_2_3}. Such a model can be engineered with an optical lattice populated by  multicomponent ultracold atoms with specific state-dependent hoppings. We show that such a setup presents a rich playground  where the low-energy excitations can be described by massless relativistic particles with arbitrary spin. Interestingly these fermions are  governed by the Weyl-like Hamiltonian $H=v_F S \cdot p$ , which is the massless version of the Dirac Hamiltonian but can have any spin, and whose eigenstates we refer to as {\it Dirac-Weyl fermions} in this paper.

More explicitly, we demonstrate that a spin-dependent hopping, tuned according to the $(2s+1)$-dimensional representations of the $\mathfrak{su}$(2)
Lie algebra, directly leads to a  regime where  the low-energy excitations are Dirac-Weyl fermions with pseudospin $s$. We show that this platform hosts two different phases at half-filling: a semimetallic phase that occurs for half-integer $s$, and a metallic phase that contains a flat zero-energy band at integer $s$. 
In the semimetallic phase, we show that a Dirac-Weyl fermion with high spin can be described as a collection of  spin 1/2 counterparts, where each species has a different effective speed of light. Accordingly, the low-energy transport is characterized by spin 1/2 Dirac-Weyl fermions moving at different velocities, an effect known as {\it multi-refringence} in the field of optics \cite{multirefringence}. As a consequence, we find exotic tunneling properties  across 
a potential barrier, which shows a remarkable multi-refringent Klein paradox.  Additionally, we study the properties of this exotic Fermi gas in the presence of a synthetic magnetic field (see~\cite{review_synthetic_gauge_fields} and references therein). We find that the 
Weyl-Landau problem can be mapped onto a {\it Dicke Hamiltonian}~\cite{dicke_model}, a well-known model in  quantum 
optics which describes the interaction of an ensemble of two-level atoms with a quantized mode of the electromagnetic radiation. From this insightful mapping, we derive the exact solution of the Hamiltonian, and predict the interesting consequences of an anomalous half-integer quantum Hall effect~\cite{anomalous_qhe}. These predictions are confirmed
by the numerical evaluation of the topological Chern numbers~\cite{cherns} and edge-states~\cite{hatsugai:2006}, which directly give the quantum Hall sequence. We obtain a general rule describing the quantum Hall effect for Dirac-Weyl fermions with arbitrary spin structures, generalizing the anomalous quantum Hall effect for pseudospin-$\frac{1}{2}$ in graphene~\cite{graphene_rev,hatsugai:2006}.

The paper is organized as follows: in Sec.~\ref{The model}, we describe the optical lattice setup to simulate the Dirac-Weyl fermions with arbitrary spin. In Sec.~\ref{spectrum}, we show how to describe a single Dirac-Weyl fermion with high spin as a collection of two-dimensional massless Dirac fermions, and also provide topological invariants that can be assigned to these excitations. In Sec.~\ref{WLLs and AQHE}, we present the anomalous quantum Hall effect which appears when these exotic particles are subjected to a synthetic magnetic field. In Sec.~\ref{klein_section}, we explore the transmission properties of these excitations in the presence of a potential barrier, leading to multirefringent Klein tunneling. In Sec.~\ref{detection}, we discuss how to probe various phenomena predicted in the main text. We present our conclusions in Sec.~\ref{Conclusions}, and leave some technical aspects to the  Appendices. In Appendix~\ref{appendix_exp}, we describe in detail how the model Hamiltonian can be simulated using laser-assisted hopping in optical superlattices. In Appendix~\ref{appendix_weyl}, we discuss the analytical solution of the model, and the calculation of the topological invariants which characterize the emerging Dirac-Weyl fermions with arbitrary spin. In Appendix~\ref{appendix_LL}, we present the details on the calculation of the Weyl-Landau levels, the zero-energy modes, and their relation to a topological index theorem. In Appendix~\ref{appendixnath}, we show the impact of a flat band on the zero-energy modes. Finally, the technical aspects of the multirefringent Klein tunneling are presented in Appendix~\ref{appendix_klein}.


\section{The model}
\label{The model}

Let us consider an ultracold Fermi gas of  $^{40}$K atoms trapped in the periodic pattern of an optical superlattice~\cite{superlattice_scheme}. The ground state of this atomic gas corresponds to the hyperfine manifold with total angular momentum $F=9/2$, and corresponding  Zeeman sublevels $m_F\in\{-9/2,...,9/2\}$. In this work, we shall  focus on a subset of $N=(2s+1)$ internal states, which will be referred to as {\it spins}, and represented by the fermionic creation (annihilation) operators $c_{\boldsymbol{r}\tau}^{\dagger}(c_{\boldsymbol{r}\tau})$. Here, $\boldsymbol{r}$ stands for the sites of a two-dimensional square superlattice, and $\tau\in\{1...N\}$ is the internal index. 
As described in Appendix~\ref{appendix_exp}, we assume that the depth of the superlattice is so large that the hopping of the atoms due to kinetic energy is inhibited. Instead this tunneling shall be laser assisted by  Raman transitions to certain excited states that are trapped in the secondary minima of the superlattice~\cite{superlattice_scheme}.  Accordingly, it is possible to control externally the spin-dependent hopping of the atoms along the primary minima of the superlattice. In 
the non-interacting limit, which can be accessed by means of Feshbach resonances, the Hamiltonian reads
\begin{equation}
\label{hamiltonian}
H=-\sum_{\boldsymbol{r},\boldsymbol{\nu}}\sum_{\tau\tau'}t_{\nu}[\mathbb{T}_{\nu}]_{\tau'\tau}c_{\boldsymbol{r}+{\boldsymbol{\nu}},\tau'}^{\dagger}c_{\boldsymbol{r}\tau}+\text{H.c.},                           
\end{equation}                                                                               
where the hopping amplitude along the $\nu=x,y$ direction is modified by a spin-dependent operator 
$\mathbb{T}_{\nu}$, and we have set the lattice spacing to $a=1$.  In this work, we will study a regime where the hopping operators correspond to a  $N$-dimensional representation of the $\mathfrak{su}(2)$ Lie algebra, namely, 
$\mathbb{T}_x= S_x$, and $\mathbb{T}_y= S_y$, which fulfill the corresponding algebra $[S_z,S_{\pm}]=\pm S_{\pm}$ and $[S_+,S_-]=2 S_z$, where 
$S_{\pm}=S_x\pm\ii S_y$.  We note that this particular Hamiltonian cannot be obtained from an external non-Abelian gauge field~\cite{non_ab_gauge_fields_lattice,dirac_fermions_2_3}, which would require that $\mathbb{T}_{x,y}$ belong to a Lie group rather than to a Lie algebra. In this paper we only consider a two-dimensional optical lattice, but the setup can also be implemented in a three-dimensional one by adding $\mathbb{T}_z= S_z$ in the third direction~\cite{superlattice_scheme}.

\section{ Dirac-Weyl fermions with arbitrary spin and flat bands }

\label{spectrum}

 In this section, we show that the low-energy excitations of the $\mathfrak{su}(2)$ Hamiltonian in Eq.~\eqref{hamiltonian} correspond to a particular type of relativistic particles, the so-called  {\it Dirac-Weyl fermions with arbitrary spin}~\cite{greiner_book}. Surprisingly, the spin-dependent hopping modifies the non-relativistic theory that describes the gas of ultracold atoms, and gives rise to emergent Dirac-Weyl fermions of arbitrary spin $s$ at low energy. In addition, we show that for half-integer spin, each Dirac-Weyl fermion with high spin can be decomposed into a collection of $N_{\text{D}}=s+\half$ independent {\it 2+1} spin 1/2 Dirac-Weyl fermions. We note that this type of excitations have raised great interest in the context of graphene~\cite{klein_tunneling}. On the other hand, for integer spin, each Dirac-Weyl fermion with high spin gives rise to $N_{\text{D}}=s$ spin 1/2 Dirac-Weyl fermions together with a single {\it zero-energy flat band}.  Recently, the physics of flat bands has also received considerable attention~\cite{weyl_1,shen:2010, apaja:2010}. The atomic system discussed here can be used to explore a variety of interesting effects at the forefront of condensed-matter and high-energy physics. 
  
\begin{figure}
	\centering
	\includegraphics[width=0.65\columnwidth]{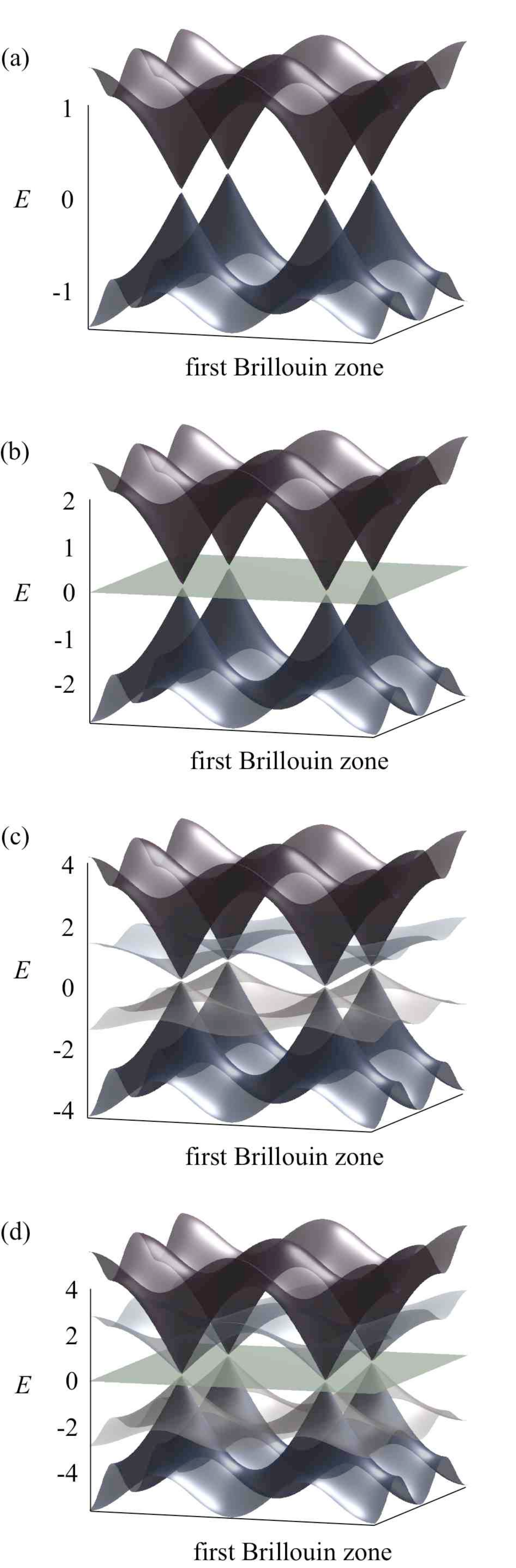}
	\caption{\label{cones} {\it Energy bands and Dirac-Weyl fermions}. Energy spectrum $E=E(k_x,k_y)$ for (a) $N$=2 (b) $N$=3 (c) $N$=4 (d) $N$=5. The wave-vector $\boldsymbol{k}=(k_x,k_y)$ belong to the first Brillouin zone.}
\end{figure}

 Transforming the Hamiltonian in Eq.~\eqref{hamiltonian} to momentum space, $c_{\boldsymbol{r}\tau}=\frac{1}{\sqrt{L}}\sum_{\boldsymbol{k}}\ee^{-\ii\boldsymbol{k}\boldsymbol{r}}c_{\boldsymbol{k}\tau}$, where $L$ is the number of lattice sites and $\boldsymbol{k}\in[-\pi,\pi)\times[-\pi,\pi)$ lies within the first Brillouin zone, one obtains the following $N$-band model 
 \begin{equation}
 \label{band_model}
 H=\sum_{\boldsymbol{k}}\Psi^{\dagger}(\boldsymbol{k})H(\boldsymbol{k})\Psi(\boldsymbol{k}), \hspace{2ex} H(\boldsymbol{k})=-\sum_{\nu}2t_{\nu}S_{\nu}\cos(k_{\nu}),
 \end{equation} 
 where the spinor $\Psi(\boldsymbol{k})=(c_{\boldsymbol{k}1},...,c_{\boldsymbol{k}N})^{t}$ contains the fermionic operators. Using the properties of the $\mathfrak{su}(2)$ Lie algebra, this Hamiltonian can be readily diagonalized (see Appendix~\ref{appendix_weyl}), leading to the following spectrum
 \begin{equation}
 \label{bands}
 E_m(\boldsymbol{k})=m\epsilon(\boldsymbol{k})=m\sqrt{(2t_x\cos k_x)^2+(2t_y\cos k_y)^2}, 
 \end{equation}
where $ m=-s,...,s-1,s$. We show in Fig. \ref{cones} the resulting band structures with $N=2, 3, 
4, 5$ internal components. Further band structures, corresponding to configurations using more internal atomic states ($N>5$), are easily derived 
from this figure. As can be observed, for $N=2$, one recovers the familiar Dirac cones that arise in graphene~\cite{klein_tunneling}. Conversely, for $N=3$ the Dirac cones are accompanied by a zero-energy flat band~\cite{weyl_1}. When $N=4$ and $N=5$, we get very interesting double-layer cones 
that host excitations with pseudospin 3/2 and 2 respectively (cf. below). As shown in Fig. \ref{cones} (c)-(d), these double-layer cones correspond to different effective speeds of light, therefore leading to the birefringence phenomenon. 

\begin{figure}[!hbp]
	\centering
	\includegraphics[width=0.5\columnwidth]{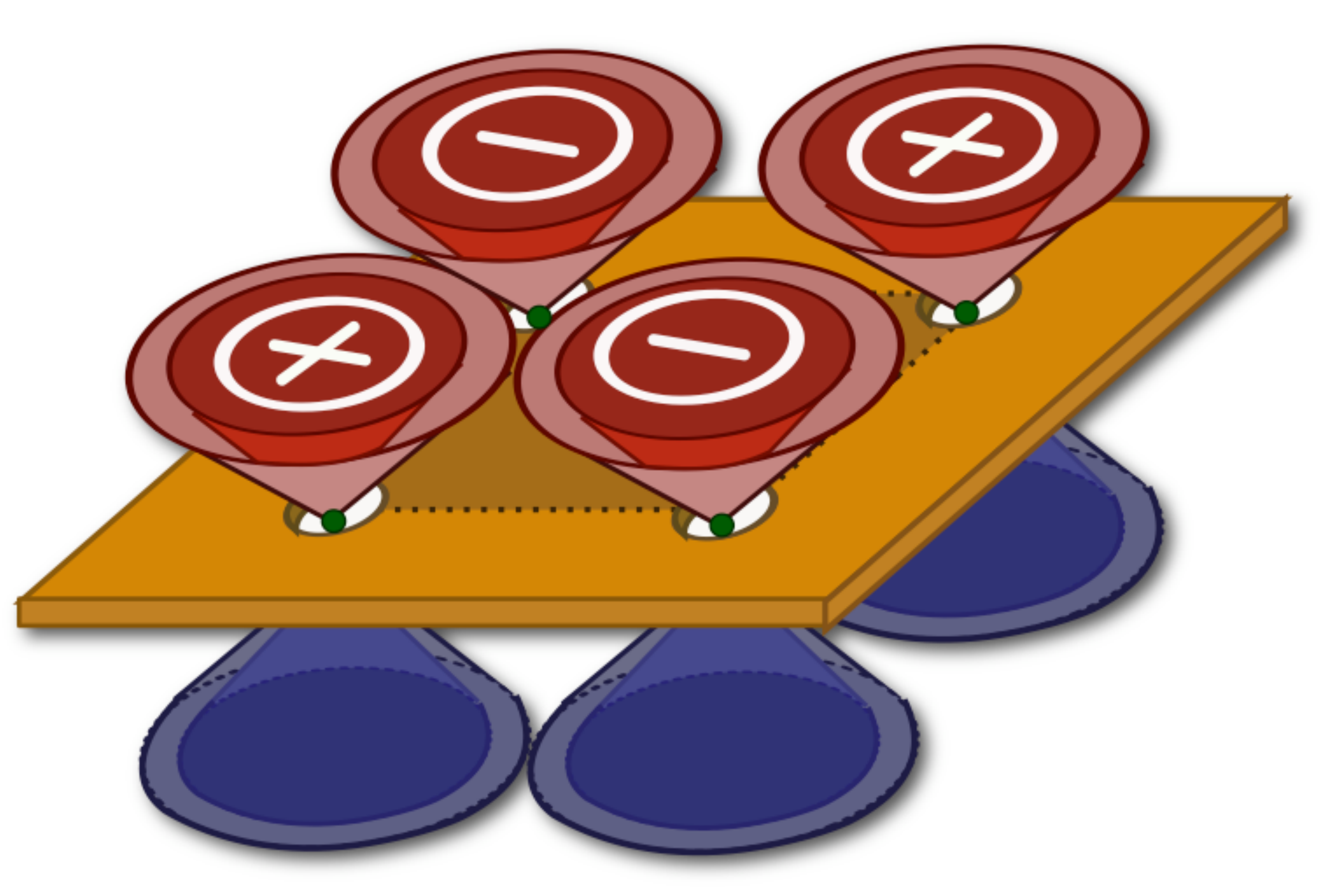}
	\caption{{\it Schematic representation of the Dirac points and their topological description}. Inside the first Brillouin zone, we find $N_{\text{W}}=4$ isolated points where conical intersections occur. Around these points, one finds that the emergent description is that of Dirac-Weyl fermions with arbitrary spin. For semi-integer spin $s$, these Dirac-Weyl fermions have an associated Berry phase of abnormal parity $\gamma^{\boldsymbol{d}}_{m}=\pi \text{mod} (2\pi)$, and an integer-valued topological charge $\mu^{\boldsymbol{d}}=\pm N_{\text{D}}\in\mathbb{Z}$, which ensure the robustness of this phase against external perturbations.  In this figure, we show the double-layered cone structure of $s=3/2$ Dirac-Weyl fermions. }
	\label{weyl_points}
\end{figure}

In general, we may conclude that  $N$ energy bands  touch at four highly-symmetric momenta $\boldsymbol{K}_{\text{W}}^{\boldsymbol{d}}=(\frac{\pi}{2}d_x,\frac{\pi}{2}d_y)$, where $d_{\nu}\in\mathbb{Z}_2$, which we shall refer to as  {\it Dirac  points} (see Fig.~\ref{weyl_points}). At half-filling,  the low-energy excitations can be described by the following Hamiltonian
\begin{equation}
\label{weyl_ham} 
 H_{\text{eff}}=\sum_{\boldsymbol{d},\boldsymbol{p}}\Psi^{\dagger}(\boldsymbol{p})H^{\boldsymbol{d}}_{\text{W}}(\boldsymbol{p})\Psi(\boldsymbol{p}),\hspace{1ex} H_{\text{W}}^{\boldsymbol{d}}(\boldsymbol{p})=\sum_{\nu}c_{\nu}d_{\nu} S_{\nu}p_{\nu},
\end{equation}
where $\boldsymbol{p}=\hbar(\boldsymbol{k}-\boldsymbol{K}^{\boldsymbol{d}}_{\text{W}})$ is the momentum around each Dirac point, and we define $c_{\nu}=2t_{\nu}/\hbar$. We stress that in the isotropic regime $c_x=c_y$, the effective  Hamiltonian for the excitations around $\boldsymbol{K}_{\text{W}}=(\frac{\pi}{2},\frac{\pi}{2})$ corresponds to that of the usual Dirac-Weyl fermions. One finds  $H_{\text{W}}=cp \Lambda_s$, where we have used the helicity operator $\Lambda_s=\boldsymbol{S}\cdot\frac{\boldsymbol{p}}{p}$, i.e. the projection of the spin along the direction of the 
linear momentum~\cite{greiner_book}. Therefore, the excitations described by Eq.~\eqref{weyl_ham} can be interpreted as $N_{\text{W}}=4$ relativistic Dirac-Weyl fermions with arbitrary spin in an underlying anisotropic spacetime with an effective speed of light $c_x\neq c_y$. We stress that for $N>2$, the $S_{\nu}$ matrices do not satisfy a Clifford algebra and therefore, the Hamiltonian in Eq. \eqref{weyl_ham}  does not correspond to a Dirac Hamiltonian. We also note that due to the fermion doubling problem~\cite{fermion_doubling}, the helicity operator cannot be globally incorporated into the lattice. \\

At this point it is worth emphasizing that in a high-energy context, Dirac-Weyl fermions are usually associated to spin-1/2 particles. Interestingly, our experimental platform allows us to synthesize Dirac-Weyl fermions with any arbitrary spin, integer or half-integer. In fact, we find two different phases that depend upon the spin $s$, and can be thus controlled experimentally. \\

{\it a) Semimetallic phases:} For semi-integer spin, $m\neq 0$ and 
the Fermi surface consists of $N_{\text{W}}=4$ isolated  points (see Fig.~\ref{weyl_points}). It is possible to find an appropriate basis (see Appendix~\ref{appendix_weyl}), where each Dirac-Weyl fermion with high spin is described by means of $N_{\text{D}}= s+\half$  spin 1/2 counterparts in 2+1 dimensions. The corresponding Hamiltonian reads
\begin{equation}
\label{dirac_ham}
H_{\text{eff}}=\sum_{\boldsymbol{d},\boldsymbol{p}}\sum_{m>0}\tilde{\Psi}_{m}^{\dagger}(\boldsymbol{p})\left(c_{xm}\alpha^{\boldsymbol{d}}_xp_x+c_{ym}\alpha^{\boldsymbol{d}}_yp_y\right)\tilde{\Psi}_{m}(\boldsymbol{p}),
\end{equation}
where $c_{\nu m}=2 t_{\nu}m/\hbar$ is the effective speed of light, $\alpha_{\nu}^{\boldsymbol{d}}=d_{\nu}\tilde{\sigma}^{\nu}_m$ are the so-called Dirac matrices in 2+1 expressed in terms of the more usual Pauli matrices, and $\tilde{\Psi}_m(\boldsymbol{p})$ is a two-component Dirac-Weyl spinor (see Appendix~\ref{appendix_weyl} for the details).  This is the Hamiltonian for spin 1/2 massless Dirac fermions .  Let us stress that in this semimetallic phase, each Dirac-Weyl fermion with high spin corresponds to a collection of spin 1/2 counterparts  propagating with a different speed of light (see Fig.~\ref{cones}). This will lead to interesting effects, such as the  multirefringenet Klein tunneling presented in Sec.~\ref{klein_section} below.

{\it b) Metallic flat-band phases:} For integer spin,  we find that each Dirac-Weyl fermion with high spin describes $N_{\text{D}}= s$ spin 1/2 counterparts in 2+1 dimensions (see Appendix~\ref{appendix_weyl}). Besides, in this case, there is always a $m=0$ spin component, and thus a zero-energy flat band. We note that the presence of this peculiar flat band shall completely modifies the properties of the system. Dispersionless energy bands have indeed remarkable consequences on single-particle and many-body properties  \cite{Tasaki2008,Lieb1989, Mielke1991,Bergman2008,Green2010}. In particular, for $N=3$ our model shares great similarities with the Lieb lattice \cite{shen:2010,goldmanlieb}, where the low-energy regime is also described by a spin-1 Dirac-Weyl Hamiltonian. Note that in this specific relativistic band configuration, the fermion doubling imposed by the Nielsen-Ninomiya theorem for lattice fermions is substituted by the flat band~\cite{Dagotto,shen:2010}. \\

{\it c) Topological invariants:} 
Due to the Hamiltonian particle-hole symmetry, the zero eigenvalues are necessarily doubly degenerate and could therefore lead to non-trivial Berry phases~\cite{berry_phase}. We have indeed demonstrated, in Appendix~\ref{appendix_weyl}, that each of the underlying spin 1/2 Dirac-Weyl fermions carries a Berry phase  $\gamma^{\boldsymbol{d}}_{m}=2\pi m\text{sgn} (d_xd_y)$, a property also familiar from graphene ~\cite{klein_tunneling}. Note that for half-integer spin, the Berry phase is of abnormal parity $\gamma^{\boldsymbol{d}}_{m}=\pi \text{mod} (2\pi)$, regardless of the Dirac  point or Dirac-Weyl fermion. Therefore our semimetal phase ($s$ half-integer) has a Berry phase of $\pi$ which is protected by the particle-hole symmetry. \\


Furthermore, the particle-hole symmetry also allows for a more complete topological description of the semimetallic phases. In Appendix~\ref{appendix_weyl}, we show that a particular topological invariant can be assigned to each Dirac  point. This topological index is given by an integer-valued topological charge $\mu\in\mathbb{Z}$ introduced in Ref.~\cite{top_charge}. In our case, we find that the different Dirac  points carry a topological charge  with a different sign given by $\mu^{\boldsymbol{d}}=\sum_{m>0}\mu^{\boldsymbol{d}}_{m}$, where $\mu^{\boldsymbol{d}}_m=\text{sgn}(d_xd_y)\in\mathbb{Z}_2$. Accordingly, we can give a topological  characterization of  each Dirac  point, which is $\mu^{\boldsymbol{d}}=\pm N_{\text{D}}\in\mathbb{Z}$ (see Fig.~\ref{weyl_points}).  For small perturbations that preserve the particle-hole symmetry, this semimetallic phase is topologically protected since these topological invariants cannot change. On the other hand, when the perturbation is strong enough, topological charges of opposite sign may annihilate each other, giving rise to interesting quantum phase transitions that generalize those studied in Ref.~\cite{tqpt}. 
In the presence of a weak harmonic confining potential, it has been found that  the Dirac points and the characteristic spectrum survives locally in the trap, provided the confining potential varies over a length scale much larger than the extent of a unit cell \cite{harmonic_confinement}.

\section{Weyl-Landau levels and the half-integer quantum Hall effect}
\label{WLLs and AQHE}

In this Section, we study a peculiar quantum Hall effect which occurs when the Dirac-Weyl fermions are subjected to an external synthetic magnetic field. The integer quantum Hall effect is characterized by the perfect quantization of the transverse Hall conductivity $\sigma_{H}= \nu \frac{e^2}{h}$, where $\nu$ is an integer. This fundamental phenomenon takes place in two-dimensional systems subjected to a strong magnetic field and exists in the non-interacting limit. Since the ultracold atoms are neutral, one needs to mimic the effects of an external magnetic field in order to perform a quantum simulation of the quantum Hall effect.   A simple way to introduce a magnetic field is to rotate the system where the Coriolis force in the rotating frame plays the same role as the Lorentz force on a charged particle in a uniform magnetic field. Alternatively, optically induced gauge potentials by means of laser-assisted tunneling can be used, see also Appendix A and Refs.~\cite{review_synthetic_gauge_fields,superlattice_scheme}. These techniques allow us to modify the hopping of Eq.~\eqref{hamiltonian} according to the so-called Peierl's substitution $\mathbb{T}_{\boldsymbol{\nu}}\rightarrow\mathbb{T}_{\boldsymbol{\nu}}\ee^{-\ii \frac{e}{h}\int_{\nu} \text{d}\boldsymbol{r}\boldsymbol{A}}$, where   $\boldsymbol{A}$ is a synthetic gauge potential giving rise to an effective magnetic field $\boldsymbol{B}=\nabla\times\boldsymbol{A}$. As shown in recent experiments~\cite{gauge_fields_exp}, ultracold atoms can indeed be used to investigate the effects of external gauge fields. In the following, we investigate the interplay between the Hall plateaus and the underlying spin structure. This study offers a generalization of the anomalous quantum Hall effect observed in graphene, where zero energy modes contribute in a fundamental manner. Here we demonstrate how this peculiar Hall sequence is indeed related to the number of Dirac points and zero energy modes, for the general case of arbitrary spin structures. This study is performed both for the lattice and for the continuum limit of our model. 

\begin{figure}
	\centering
	\includegraphics[width=1\columnwidth]{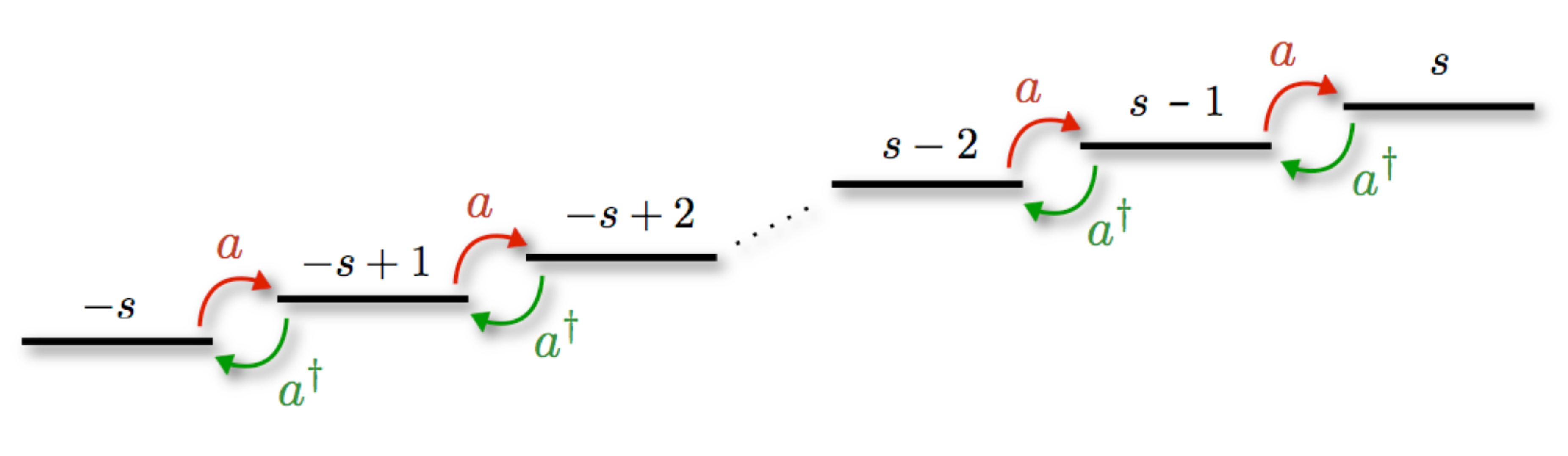}
	\caption{{\it Ladder scheme of the isotropic Weyl-Landau Hamiltonian}.  The system with  pseudospin $s$ can be understood as a $(2s+1)$-level atom, where transitions between the different spin projections $m\to m+1$ ($m\to m-1$) are dressed by the annihilation $a$ (creation $a^{\dagger}$) of a  motional quanta. }
	\label{weyl_landau}
\end{figure}

{\it a) Continuum description: Weyl-Landau levels.} For small-enough gauge fields, one can show that the  Peierl's substitution leads to the usual minimal coupling performed in the effective Hamiltonian ~\eqref{weyl_ham}.  
The standard procedure is to
replace the canonical momentum by a gauge-invariant quantity $\boldsymbol{p}\rightarrow\boldsymbol{\Pi}=\boldsymbol{p}+e\boldsymbol{A}$, whose components no longer commute 
 $[\Pi_x,\Pi_y]=-\ii \hbar^2/l_{\text{B}}^2$, where $l_{\text{B}}=\sqrt{\hbar/eB}$ is the magnetic length. By introducing the bosonic creation-annihilation operators $a^{\dagger}=\frac{l_{\text{B}}}{\sqrt{2}\hbar}(\Pi_x+\ii\Pi_y)$ where $a=(a^{\dagger})^{\dagger}$, the effective Weyl-like Hamiltonian in Eq.~\eqref{weyl_ham} is recast into
\begin{equation}
\label{dicke_model}
H_{\text{WL}}^{\boldsymbol{d}}=H_{\text{W}}^{\boldsymbol{d}}(\boldsymbol{p}+e\boldsymbol{A})=g_{\boldsymbol{d}+}aS_++g_{\boldsymbol{d}-}aS_-+\text{H.c.}
\end{equation}

where we have introduced $g_{\boldsymbol{d}\pm}=\hbar(c_xd_x\pm c_yd_y)/(2\sqrt{2}l_{\text{B}})$.
Interestingly the problem of Dirac-Weyl fermions subjected to a magnetic field~\cite{comment}, hereafter referred to as the {\it Weyl-Landau levels} (WLLs),  can be exactly mapped onto the so-called {\it Dicke model} which is well known from quantum optics~\cite{dicke_model}. This model, which describes the interaction between a collection of two-level atoms and a single mode of the quantized electromagnetic field, displays a wide range of interesting phenomena (see e.g.~\cite{dicke_rev}). In Fig.~\ref{weyl_landau}, we represent schematically the physical content of the Weyl-Landau Hamiltonian of Eq.~\eqref{dicke_model} in the isotropic regime, namely  $c_x=c_y=:c$, and for $\boldsymbol{d}=(1,1)$. In this situation, $g_{\boldsymbol{d}-}=0$, $g_{\boldsymbol{d}+}=\hbar c/\sqrt{2}l_{\text{B}}=:g$,  and the spin raising (lowering) transitions are dressed by the annihilation (creation) of motional bosonic quanta.    

In Appendix~\ref{appendix_LL}, we present a compact way to solve the Weyl-Landau Hamiltonian in Eq.~\eqref{dicke_model} for $\boldsymbol{d}=(1,1)$, in the isotropic regime $c_x=c_y$. This solution is iterative in nature and
allows us to take the analytical expressions of the WLLs from pseudospin $s$ to $(s+\half)$. From this method, one can derive the exact energy spectrum 
for different pseudospins, such as $s=\{\half,1,\frac{3}{2},2\}$ presented in Table~\ref{WLL_spectrum}. In Table~\ref{WLL_spectrum}, we also find  that the energy spectrum of the $s=\half$ WLL is analogous to the relativistic Landau levels in graphene~\cite{graphene_rev}. We also observe the characteristic dependence of the energies, $E\propto g\propto\sqrt{B}$, which is a hallmark that guarantees the relativistic nature of the particles. As can be seen in the Table~\ref{WLL_spectrum}, this peculiar dependence, $E\propto\sqrt{B}$, is also fulfilled in higher pseudospin cases. In the $s=1$ case, we observe a couple of particle-hole symmetric levels with analogous properties, but also a novel zero-energy Landau level which is completely absent in the half-integer spin case. The presence of this particular zero-energy WLL will have important consequences in the quantum Hall response of the system. Finally, for $s=\frac{3}{2}$ and $s=2$, we observe two pairs of particle-hole symmetric levels, which is a  consequence of the two underlying species of  spin 1/2 Dirac-Weyl fermions  assigned to each Dirac  point. Note also that the dependence on the number of motional quanta $n$ gets more involved as the pseudospin is increased.

\begin{table}

		\begin{tabular}{l  l  l  l}
		\hline 
		\hline

		 $s$ & Weyl-Landau levels $E_{nj}$  &\hspace{-1ex} Constraint \hspace{-1ex}\\

		\hline
		\hline

			\vspace{-1ex}\\
		$\half$ & $E_{n,\pm}=\pm g\sqrt{2n}$ & $\hspace{2ex}n\geq1$\\
		
		\vspace{-1ex}\\
		
		\hline

			\vspace{-1ex}\\
			
			$1$ & $E_{1,\pm}=\pm g\sqrt{2}$ & $\hspace{2ex}n=1$ \\
	          & $E_{n,\pm}=\pm g\sqrt{2(2n-1)} $ & $\hspace{2ex}n\geq2$\\
		        & $E_{n,0}=0$ & \\
		        
		            	\vspace{-1ex}\\
		            
		 \hline
		 
		 	\vspace{-1ex}\\
		 
		 $\frac{3}{2}$ & $E_{1,\pm}=\pm g\sqrt{3}$ & $\hspace{2ex}n=1$\\
		                 & $E_{2,\pm}=\pm g\sqrt{10}$ & $\hspace{2ex}n=2$\\
		                 & $E_{n,\pm,1}=\pm g\sqrt{(5n-1)-\sqrt{16(n-1)^2+9}}$ & $\hspace{2ex}n\geq3$\\
		                 & $E_{n,\pm,2}=\pm g\sqrt{(5n-1)+\sqrt{16(n-1)^2+9}}$ & \\

		            	\vspace{-1ex}\\
		  \hline
		  
		  	\vspace{-1ex}\\
		  
		 $2$  & $E_{1,\pm}=\pm 2g $ & $\hspace{2ex}n=1$\\
		             & $E_{2,\pm}=\pm g\sqrt{14}$ & $\hspace{2ex}n=2$\\
		             & $E_{3,\pm,1}=\pm g\sqrt{15-3\sqrt{17}}$, $\hspace{1ex}E_{3,\pm,2}=\pm g\sqrt{15+3\sqrt{17}}$ & $\hspace{2ex}n=3$\\
		             & $E_{n,\pm,1}=\pm g\sqrt{5(2n-3)-3\sqrt{4n^2-12n+17}}$ & $\hspace{2ex}n\geq4$\\
		             & $E_{n,\pm,2}=\pm g\sqrt{5(2n-3)+3\sqrt{4n^2-12n+17}}$ & \\
		             & $E_{n,0}=0$ & \\

		             	\vspace{-1ex}\\

		             \hline
		             \hline 

		\end{tabular}
		
		\caption{ {\it Analytical expression of the Weyl-Landau levels}. The energies corresponding to the Landau levels of Dirac-Weyl fermions with different pseudospin  $s=\{\half,1,\frac{3}{2},2\}$ are expressed as a function of the coupling strength $g=\hbar c/\sqrt{2}l_{\text{B}}$, and the number of motional bosonic quanta $n$. }
		\label{WLL_spectrum}
\end{table}

In addition to the WLLs presented in Table~\ref{WLL_spectrum}, we also study the presence of certain  special topological solutions that occur at zero energy (see the details in Appendix~\ref{appendix_LL}). These solutions, the so-called {\it zero-energy modes}, play a key role in the quantum Hall response of the sample and  give rise to the half-integer anomaly~\cite{anomalous_qhe}. The underlying topological modes contribute with a fractional transverse conductivity (in units of $e^2/h$), even in the absence of interactions. In Table~\ref{WLL_zero_modes}, we show that Dirac-Weyl fermions with a half-integer spin $s$ support $N_{\text{z}}=s+\half$ topological zero-modes which are protected by a topological Atiyah-Singer index theorem~\cite{atiyah_index,aharonov_casher}.  Besides, they present half the degeneracy of higher Landau levels, and thus lead to a half-integer anomaly in the quantum Hall response.  On the other hand, for integer spin $s$, there are also nontopological zero-energy Landau levels that arise from the highly degenerate flat band (see Table.~\ref{WLL_spectrum}). As argued in Appendix~\ref{appendix_LL}, these zero modes are not related to an index theorem and thus, are not protected.  This highly-degenerate zero-energy band characterizing the integer-spin case is responsible for the vanishing of  the half-integer anomaly. As confirmed numerically in subsection {\it b)} (cf. below), the Hall conductivity of the system fulfills
\begin{equation}
\label{qhe}
\begin{split}
\sigma_{xy}&=\frac{e^2}{h} N_{\text{W}}\left( \nu+\frac{N_{\text{z}}}{2}\right), \hspace{2ex} s\hspace{1ex}\text{half-integer}, \\
\sigma_{xy}&=\frac{e^2}{h}  N_{\text{W}}\nu, \hspace{11ex} s\hspace{1ex}\text{integer},
\end{split}
\end{equation}
where $\nu=0,1,...$ determines the different plateaus, and $N_{\text{W}}=4$ is the number of Dirac points. Therefore, as stated above, the half-integer anomaly is only valid for the half-integer spin Dirac-Weyl fermion. It is also important to note that due to the fermion doubling~\cite{fermion_doubling}, $N_{\text{W}}=4$ in our case, the fractional character of the Hall sequence is lost.

\begin{table}

		\begin{tabular}{l  l  l  l}
		\hline 
		\hline

		 $s$ & 1$^{\text{st}}$ zero mode (n=0)  & 2$^{\text{nd}}$ zero mode (n=2) \\

		\hline
		\hline

			\vspace{-1ex}\\
		$s=\half$ & $\ket{E_0}=\ket{\half,-\half}\ket{0}$ &\\
		
		\vspace{-1ex}\\
		
		\hline
		
	         \vspace{-1ex}\\
		$s=1$ & $\ket{E_0}=\ket{1,-1}\ket{0}$& \\
		
		\vspace{-1ex}\\
		
		\hline
		
	         \vspace{-1ex}\\
		$s=\frac{3}{2}$ & $\ket{E^{(1)}_0}=\ket{\frac{3}{2},-\frac{3}{2}}\ket{0}$ & $\ket{E^{(2)}_0}=\sqrt{\frac{2}{5}}\ket{\frac{3}{2},-\frac{3}{2}}\ket{2}-\sqrt{\frac{3}{5}}\ket{\frac{3}{2},\frac{1}{2}}\ket{0}$\\

		\vspace{-1ex}\\
		
				\hline
		
		         \vspace{1ex}\\
		
		$s=2$ & $\ket{E^{(1)}_0}=\ket{2,-2}\ket{0}$ & $\ket{E^{(2)}_0}=\sqrt{\frac{3}{7}}\ket{2,-2}\ket{2}-\sqrt{\frac{4}{7}}\ket{2,0}\ket{0}$\\

		\vspace{-1ex}\\

		             \hline
		             \hline 

		\end{tabular}
		
		\caption{ {\it Analytical expression of the zero-energy modes}. The Weyl-Landau Hamiltonain in Eq.~\eqref{dicke_model} also yields certain zero-energy modes whenever the constraints over the number of motional quanta presented in Table~\ref{WLL_spectrum} are not fulfilled. In this Table, we list these topological zero-energy modes for pseudospins  $s=\{\half,1,\frac{3}{2},2\}$, where $\ket{s,m}$ refers to the spin state, and $\ket{n}$ to the motional Fock state.  }
		\label{WLL_zero_modes}
\end{table}

{\it b) Lattice description: Computing the Chern numbers.} The Hall conductivity can be evaluated numerically by diagonalizing the full tight-binding Hamiltonian in Eq. \eqref{hamiltonian} after the Peierl's substitution $\mathbb{T}_{\boldsymbol{\nu}}\rightarrow\mathbb{T}_{\boldsymbol{\nu}}\ee^{-\ii \frac{e}{h}\int_{\nu} \text{d}\boldsymbol{r}\boldsymbol{A}}$. Considering the standard Kubo formula, the Hall conductivity is given by the TKNN expression~\cite{tknn} as follows
\begin{align}
\sigma_H&=\frac{e^2}{h} \sum_{E_{\alpha}<E_{\text{F}}} \frac{i}{2 \pi} \int_{\mathbb{T}^2} d \boldsymbol{k} \, \langle \frac{\partial u_{\alpha}}{\partial k_{x}}  \vert \frac{\partial u_{\alpha}}{\partial k_{y}  }  \rangle -\langle \frac{\partial u_{\alpha}}{\partial k_{y}}  \vert \frac{\partial u_{\alpha}}{\partial k_{x}  }  \rangle \\
&= \frac{e^2}{h} \sum_{\rm occ. \, bands } N_{\text{ch}} (\text{band}) ,
\label{hall}
\end{align}
where $\ket{u_{\alpha}}$ are single-particle eigenstates of \eqref{hamiltonian} and  where the Chern numbers associated to each occupied band, $N_{\text{ch}} (\text{band})$, can be efficiently computed using the method of Ref. \cite{cherns}. Here $E_{\text{F}}$ denotes the Fermi energy, which can be tuned in our setup by varying the atomic filling.

\begin{figure}
	\centering
	\includegraphics[width=0.8\columnwidth]{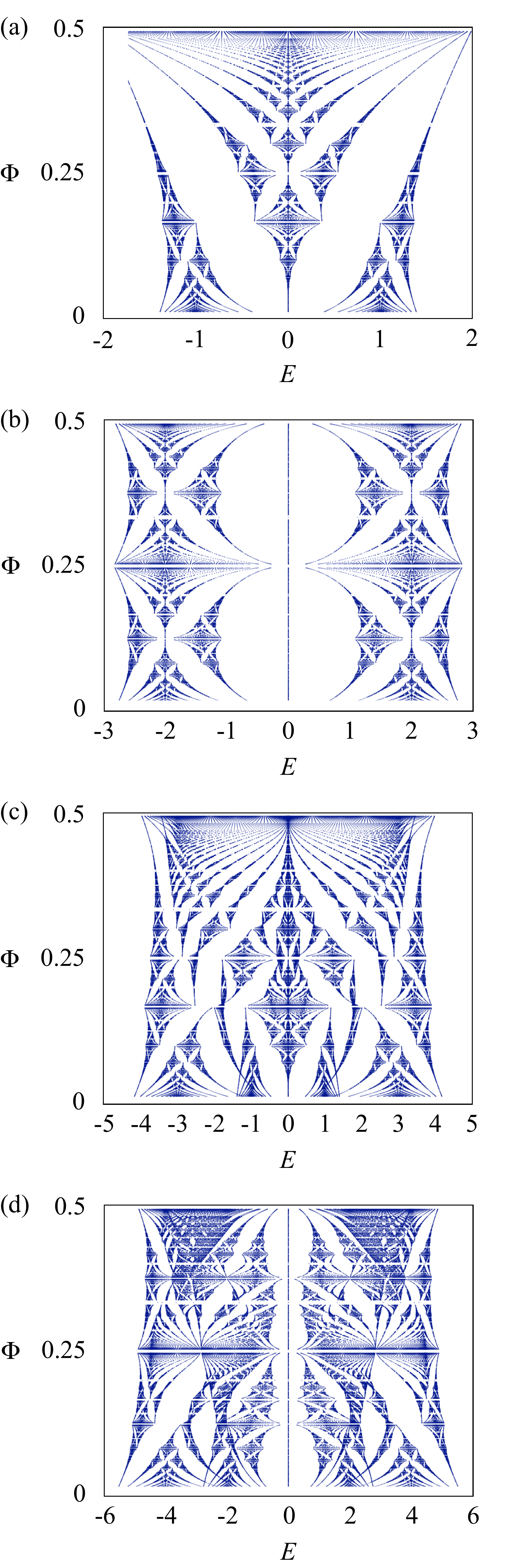}
	\caption{\label{butterflies} {\it Hofstadter-like fractal butterflies in the Dirac-Weyl fermion system}. The energy spectrum $E=E(\Phi)$ as a function of the magnetic flux shows a fractal butterfly structure: (a) $N=2$ ($s=\half$), (b) $N=3$ ($s=1$), (c) $N=4$ ($s=\frac{3}{2}$) and (d) N=5 ($s=2$). The parabolic dependence at low magnetic fluxes can be related to the underlying relativistic fermions. Note also that for integer spin, one gets a flat band exactly located at zero energy.}
\end{figure}

Before analyzing the specific Hall plateaus of the systems associated to different spin structures, let us draw their energy spectra $E=E(\Phi)$ as a function of the dimensionless magnetic flux $\Phi=2 \pi B a^2$. These computed spectra generalize the famous Hofstadter butterfly~\cite{hofstadter}. The fractal butterfly spectra corresponding to the cases $N=2, 3, 4, 5$ are illustrated in Fig.\ref{butterflies}. For $N=2$, one recovers the spectrum of the $\pi$-flux model \cite{hatsugai:2006}. For $N=3$ one observes a spectrum similar to the Lieb lattice \cite{aoki:1996,goldmanlieb}, which highlights the similarity between these two models that share a spin-1 configuration. We note the existence of a highly-degenerate flat band lying exactly at zero energy. For $N>3$, the spectra become more complex and show complicated overlaps between butterfly-like substructures. In particular, one identifies two overlapping butterflies in Fig.~\ref{butterflies}(c)-(d), each of which belongs to one of the two species of  spin-$1/2$ Dirac-Weyl fermions for $s=3/2$ and $s=2$. We note that for integer spin ($N$ odd), the central flat band at $E=0$ remains robust for all flux $\Phi$. 

\begin{figure}
	\centering
	\includegraphics[width=1\columnwidth]{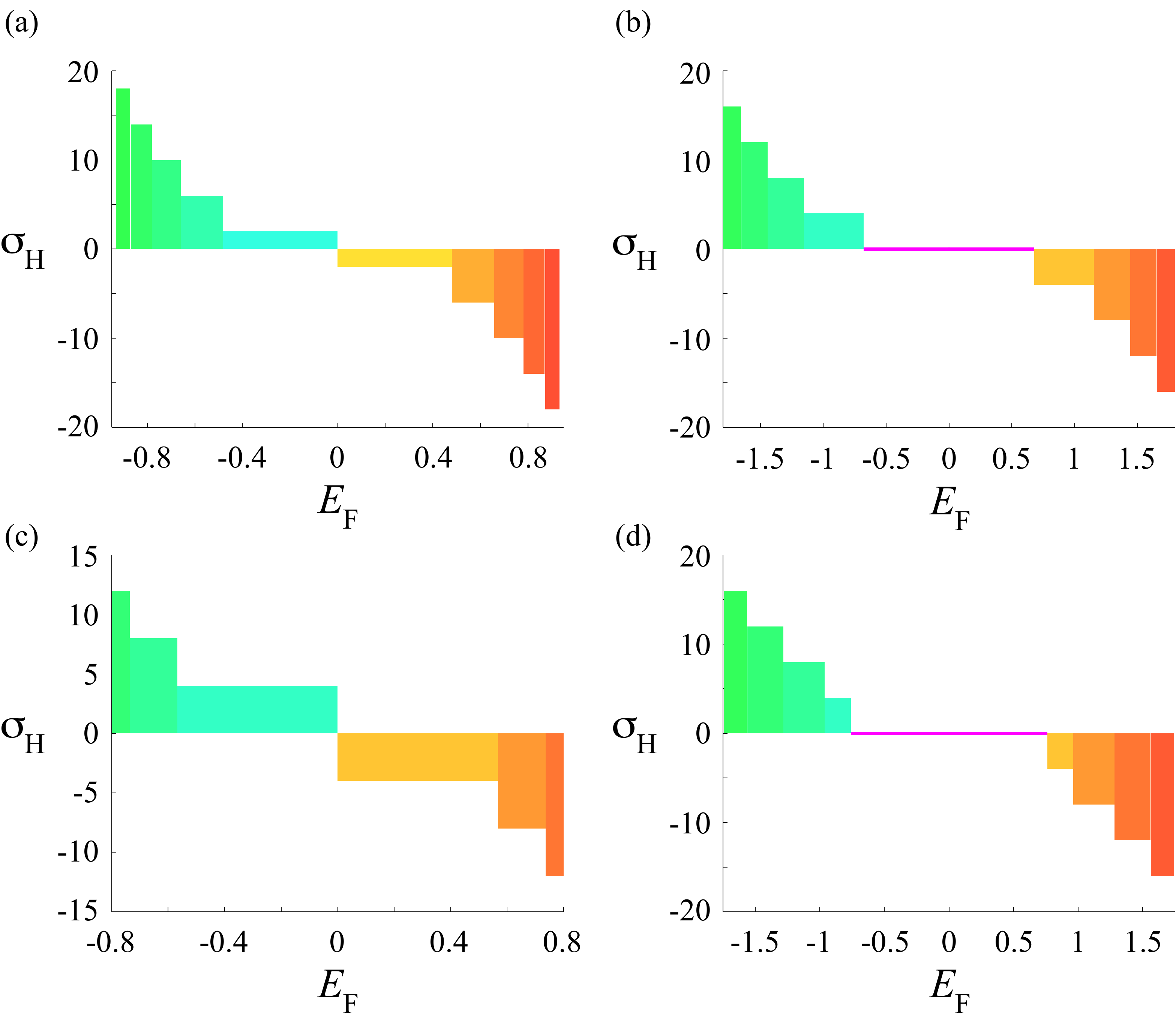}
	\caption{\label{plateaus} {\it Quantum Hall effect in the Dirac-Weyl fermion system.} The transverse Hall conductivity as a function of the Fermi energy $\sigma_{H}(E_{\text{F}})$ displays a sequence of plateaus that are associated to an underlying topological order:  (a) $N=2$ ($s=1/2$), (b) $N=3$ ($s=1$), (c) $N=4$ ($s=3/2$) and (d) N=5 ($s=2$). Here we set the magnetic flux to  $\Phi=1/51$.}
\end{figure}

The Hall plateaus corresponding to $N=2,3,4,5$ are illustrated in Fig. \ref{plateaus}. We expect the Hall sequences to be compatible with the continuum analysis in the low flux regime, and we therefore set $\Phi=1/51$ in this analysis. We also focus on the low-energy range, where the description in terms of Weyl-Landau levels is valid. First we note that for all the cases  (Fig. \ref{plateaus} (a)-(d)), steps of $N_{\text{W}}=4$ are observed in the ranges $E_{\text{F}}<0$ (hole) and $E_{\text{F}}>0$ (particles). The differences between these several Hall sequences occur at half-filling ($E_{\text{F}}=0$), where the flat band and the number of zero modes play a fundamental role. For $N=2$ (spin-$\half$), one observes a central step of $N_{\text{W}} N_{\text{Z}}=4 \times 1=4$, while for $N=4$ (spin-$\frac{3}{2}$), one gets a central step of $N_{\text{W}} N_{\text{Z}}=4 \times 2=8$. This numerical results confirm the prediction of Eq.~\eqref{qhe} based on the analytical expressions for the Dirac  points, and zero-energy modes derived earlier. 

For general $N$ even (half-integer spin), one indeed obtains that the two central plateaus, around half-filling, are given by $\sigma_{xy}=\pm N_{\text{W}} N_{\text{Z}}/2$ (in units of $e^2/h$), therefore giving a fundamental signature of the number of Dirac  points and zero modes. 

For $N$ odd (integer spin), one finds a vanishing contribution of the zero modes to the Hall conductivity: a Hall plateau corresponding to $\sigma_{xy}$=0 is clearly observed in the vicinity of $E_{\text{F}}=0$ (cf. Fig. \ref{plateaus} (b),(d)). One can understand the vanishing of the half-integer anomaly as a consequence of the lack of an index theorem for the zero modes \cite{comment2}.  Note  that this general result is in perfect agreement with the Hall sequences computed for the $\mathcal{T}_3$ \cite{weyl_1} and Lieb \cite{goldmanlieb} lattices (i.e. lattices with pseudospin 1).

\begin{figure}
	\centering
	\includegraphics[width=1\columnwidth]{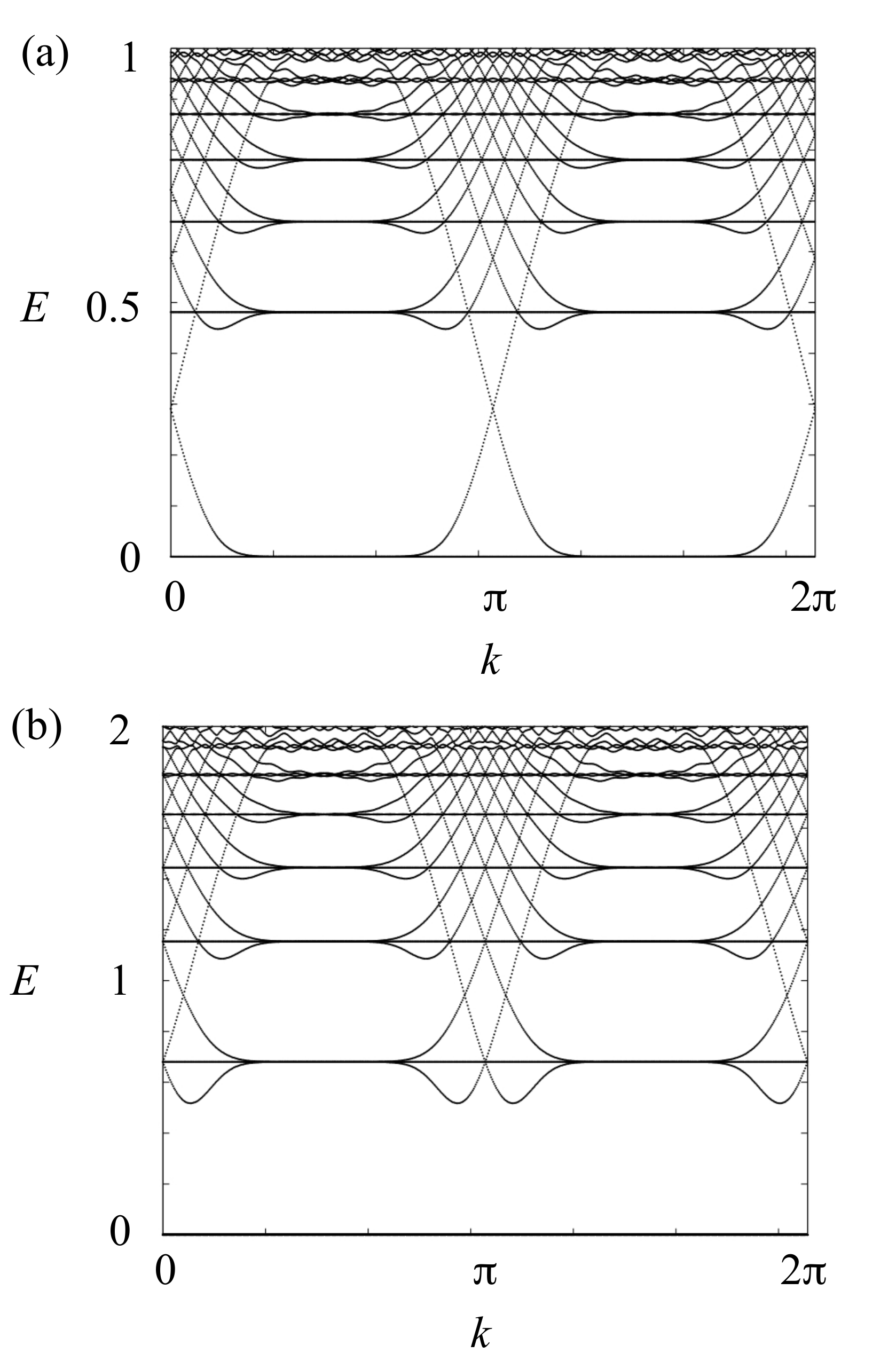}
	\caption{\label{edges} {\it Energy spectrum and current-carrying edge states.} Energy spectrum $E=E(k)$ for (a) $N$=2 and (b) $N$=3. Here we set $\Phi=1/51$. }
\end{figure}

In order to deepen our understanding of the different Hall conductivity plateaus for {\it odd} and {\it even} $N$ cases, we further investigate the associated edge-states, which can be obtained by diagonalizing the system in a cylindrical geometry. In other words, the topological properties hidden in the bulk (i.e. the Chern numbers) may be visible around the boundaries by the holographic bulk-to-edge correspondence \cite{hatsugai:2006}. This correspondence is based on the fact that the edge-states carry the Hall current along the boundaries of the system. In Fig. \ref{edges}, we show the corresponding energy spectra $E=E(k)$, where $k$ is a Bloch parameter, for $N=2$ (half-integer spin) and $N=3$ (integer spin). Note that Fig. \ref{edges} (a) is similar to the spectrum of graphene [cf. Fig. 21 in Ref. \cite{graphene_rev}], and highlights the anomalous quantum Hall effect that occurs for half-integer spins:  the contribution of the zero modes at half-filling confirms the aforementioned result $\sigma_{xy}(E_{\text{F}}=0^+)= N_{\text{W}} N_{\text{Z}}/2$ (in units of $e^2/h$). In Fig. \ref{edges} (a), we observe that each boundary is populated by two edge-states: this is due to the presence of four Dirac  points in the first Brillouin zone (in contrast with the two Dirac points of graphene that lead to a single edge-state). Fig. \ref{edges} (b) emphasizes the absence of gapless edge-states in the first gap above $E=0$, as observed for all the cases corresponding to odd $N$. This analysis confirms the general result presented in Eq. \eqref{qhe}.

The absence of an anomalous quantum Hall effect for integer $s$ is certainly interesting. As discussed above, this effect is also manifested by the zero Hall plateau around $E_{\text{F}}=0$ (cf. Fig. \ref{plateaus} (b),(d)) or by the absence of visible edge-states stemming from the zero energy modes (cf.  Fig. \ref{edges} (b)). The absence of an index theorem in this case prevents the robustness of the topological zero modes and therefore, it is  reasonable to argue that they cannot contribute to the Hall conductivity (since this physical observable is topologically protected). Furthermore, the localized properties of the states rooted in the flat band \cite{apaja:2010} could also explain that their associated zero-modes would potentially contribute to edge-states with zero velocity (i.e. they would not contribute to the Hall conductivity). The latter effect is further detailed in Appendix \ref{appendixnath}.

\section{Klein multi-refringence tunnelling}
\label{klein_section}

\begin{figure}
	\centering
	\includegraphics[width=1\columnwidth]{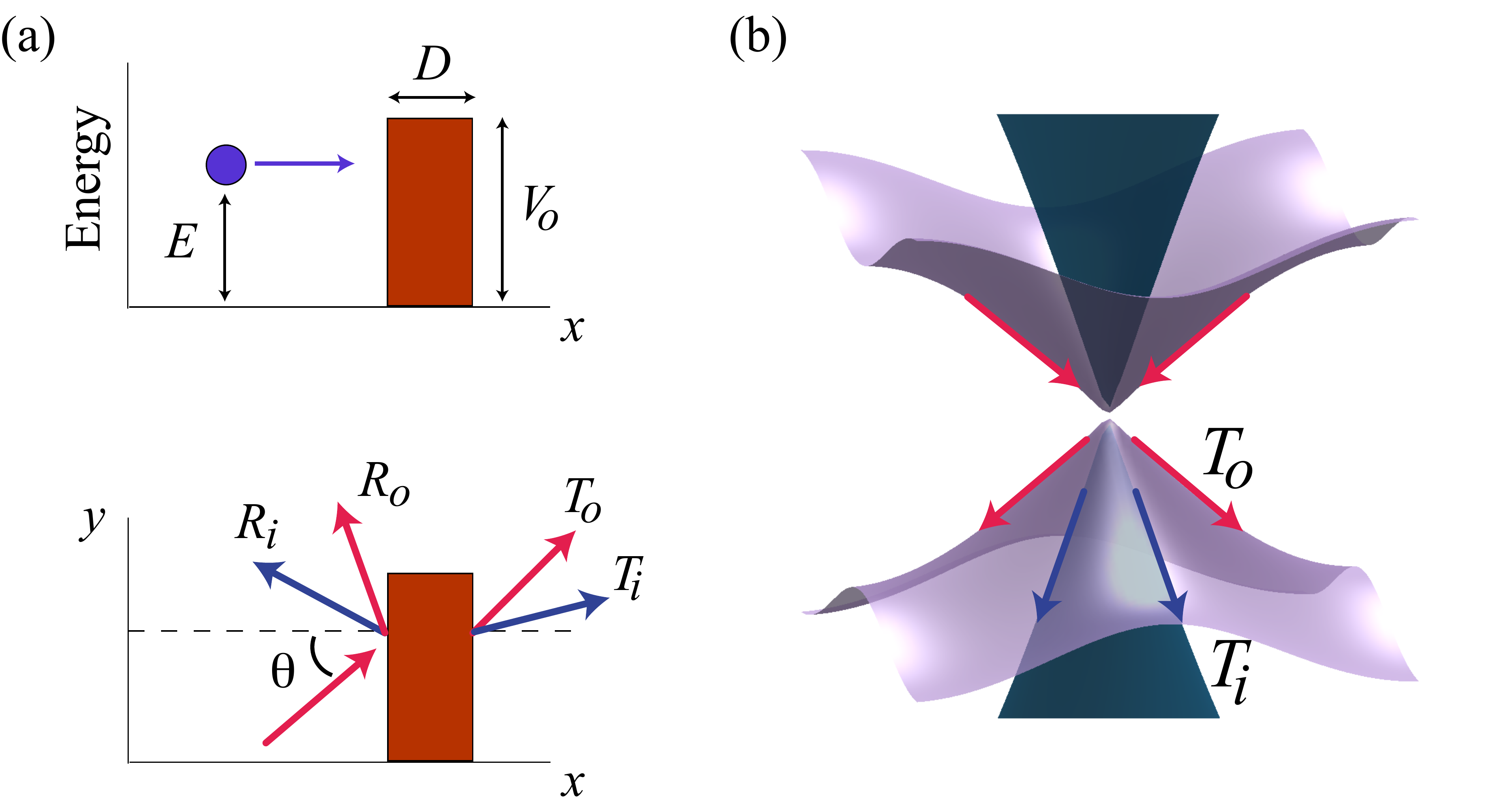}
	\caption{\label{klein1} {\it Klein birefringent tunneling}. Schematic view of a spin $3/2$ Dirac-Weyl fermion  incident on a potential barrier (a) while following the outer cone (b).}
\end{figure}

\begin{figure}
	\centering
	\includegraphics[width=1\columnwidth]{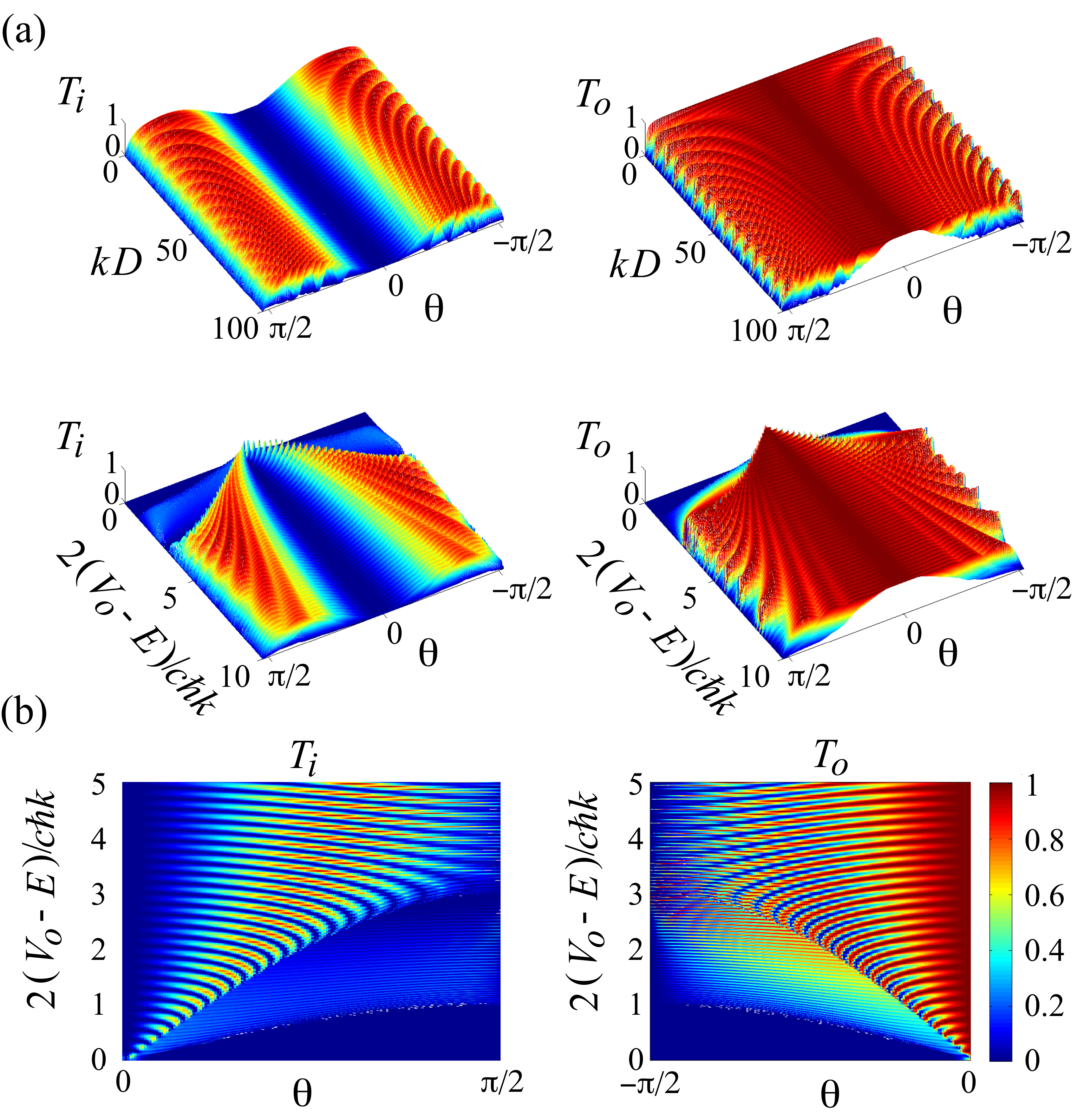}
	\caption{\label{klein2} {\it Klein birefringent tunneling}. Double transmissions (birefringence) of spin $3/2$ Dirac-Weyl particles from the double-layered cone structures as the width and height of the barrier change. In (a), the transmission diagram with the increase of width (up two panels) and height (down two panels) of potential barrier are shown. Parameters used in the simulation: up two panels, $2(V_0-E)/c\hbar k=5$ fixed with the evolution of $kD$; down two panels with $kD=50$  fixed, and the evolution of $2(V_0-E)/c\hbar k$. (b) shows the blowup of initial increase with height of the barrier of the down two panels in (a) to illustrate the role of evanescent waves, where the cutoff conditions for helicities  $h_1=1/2 $ and $h_2=3/2 $ are clearly visable. }
\end{figure}

It was shown in~\cite{klein_tunneling} that due to the coupling of positive and 
negative chirality channels outside and inside a potential barrier, quantum tunneling of Dirac 
particles in graphene becomes highly anisotropic, and the barrier remains perfectly transparent for normal incidence. The fermionic and bosonic Klein paradox are both discussed in the literature (see~\cite{Bosonic KT} and references therein). It was shown that while the central mechanism for fermionic Klein tunneling is the Pauli exclusion principle, the key to bosonic Klein tunneling is stimulated emission. Jakubsky {\it et al.}~\cite{klein_supersymmetry} were able to show that for normal incidence, 
the potential can be gauged away by a unitary transformation, leaving a free particle dynamics in disguise, where a unitary equivalence between the transformed Hamiltonians becomes encoded in a supersymmetry algebra. Evidence for Klein tunneling has been obtained from graphene pn junction~\cite{klein_graphene_pn1, klein_graphene_pn2} and most recently been simulated using trapped ions~\cite{klein_simulation_ions}.

The different helicities carried by the Dirac-Weyl fermions can naturally couple inside the barrier, but 
there is also a possibility to transform one helicity to another outside the barrier, resulting in a remarkable multirefringence phenomenon familiar from optics~\cite{multirefringence}. In Appendix~\ref{appendix_klein} we present a detailed treatment of the Klein tunneling of Dirac-Weyl fermions with arbitrary spin. Using the equations introduced there,  
a numerical investigation of the double-layered cone structure of spin $3/2$ particles shows that Klein-birefringence  is indeed present (see Fig.~\ref{klein1} for a schematic overview). The incident energy of the particle is $E$, where the barrier width is $D$ and height $V_0$ respectively. The incident particle is chosen to follow the outer layer cone. As such, the evanescent waves do not cause a cutoff for any helicity outside the barrier, see Fig.~\ref{klein1}(b).

Fig.~\ref{klein2} shows the transmission of spin $3/2$ Dirac-Weyl fermions when the width and height of the barrier changes. For normal incidence, the barrier is perfectly transparent as in graphene~\cite{klein_tunneling}, see Fig.~\ref{klein2}(a). When the incident angle deviates from normal incidence, the component following the inner cone shows periodic peaks in the transmission, which is the hallmark of birefringence. There are no resonant conditions, where the barrier would become perfectly transparent at certain incident angles, apart from normal incidence. For lower barriers, evanescent waves play an important role when coupling helicities outside and inside the barrier, as shown in the two lowest panels in Fig.~\ref{klein2}(a) and Fig.~\ref{klein2}(b). For a low enough barrier, and for an incident wave beyond the critical angle $\theta_{c1} =arcsin((V_0-E)/c\hbar kh_1), h_1=1/2$, all the helicities coupled inside the barrier are evanescent waves, hence there is no transmission for any component as shown in Fig.~\ref{klein2}(b). For a combination whose critical angle is $\theta_{c2} =arcsin((V_0-E)/c\hbar kh_2)$ and $h_2=3/2$, there is only one negative helicity of $h_1=1/2 $ inside the barrier which is coupled in the form of a propagating wave. The transmission properties are consequently modified dramatically in this regime, see Fig.~\ref{klein2}(b). These two boundaries stemming from the helicities of $h_1=1/2 $ and $h_2=3/2 $ are clearly visible in Fig.~\ref{klein2}(b). Herein lies the paradox. For low barriers, there is no transmission, while for high barriers, the transmission is high. For the case of an incident particle following the inner cone, there is an additional cutoff condition exerted by evanescent waves with helicity $h_2=3/2 $ outside the barrier, i.e.,  beyond the critical angle $\theta_{c}=arcsin(h_1/h_2)=arcsin(1/3)=0.34$, a particle with helicity $h_2=3/2$ cannot be transmitted. 

For n-layered cones we obtain similar results, where inside the barrier the presence of evanescent waves will exert a cutoff for each helicity, where with increasing barrier height transmission is gradually allowed for the different helicities. Outside the barrier, the evanescent waves only exert a cutoff when a small helicity is transformed to a large one. The transmission spectrum typically shows n-refringence, which we refer to as Klein multirefringent Tunneling. This multi-refringence is a result of the rather unique and non-trivial helicity of these Dirac-Weyl fermions.

\section{Detection}
\label{detection}

The experimental realisations of the proposed scenarios and the detection of the resulting effects are certainly challenging but should still be within experimental reach with state of the art trapping and manipulation of atomic ensembles.
Any detection scheme capable of resolving the effects discussed in this paper will typically have to be able to distinguish between spin states, particle density and momentum distribution. We briefly outline here some of the possible techniques one can use, and refer the reader to  \cite{tqpt} for an extensive discussion about detection methods of exotic quasi particles in optical lattices.

Regarding the number of Dirac  points we note that this can be addressed by measuring the atom density close to a zero chemical potential \cite{zhu2007}. To obtain the number of zero modes and the number of Dirac points one can evaluate, or indeed measure, the Hall conductivity around half-filling. This could be achieved using the Streda formula based on the atomic density which can be measured precisely by the {\it in situ} individual atom detection as in Ref. \cite{SAD1,SAD2}. In addition, a measurement of the atom density as a function of the chemical potential allows us to map the density of states (DOSs), where the Van Hove singularities in the DOSs are directly related to the number of layers of the cone structure, see Ref. \cite{dirac_fermions_2_3}. By mapping out the momentum distribution using atomic angle-resolved photoemission spectroscopy (ARPES) Ref. \cite{ARPES}, momentum-resolved Bragg spectroscopy Ref. \cite{MRBS}, or adiabatic release Ref. \cite{release}, allows us to map the fermi surface and thus the location of each Dirac  points.

The existence of a flat band with integer spin can be detected by its energy dispersion and its related wave function. The flat band gives a peak in the DOSs \cite{weyl_1}, which can be detected by measuring the atomic density.  More importantly, the localization properties resulting from the flat band could also be detected \cite{apaja:2010}. This localization can be observed after the weak harmonic confinement of the atoms is removed but with the optical lattice kept in place: the atoms occupying the Dirac cones will fly away fast while the atoms occupying the flat band will remain stuck in the immobile flat band states as shown in \cite{apaja:2010}. In our system, this localization property should be observed for $N$ odd and can be understood by observing the vanishing of the Dirac-Weyl spinor components corresponding to odd cyclotron modes (see for instance Eq.(\ref{zero_spinor}) in Appendix~\ref{appendix_LL}) -- in direct analogy with the flat band wave function of the Lieb lattice. The vanishing of the components corresponding to the odd cyclotron modes of the Dirac Weyl spinor can also be confirmed by the color resolution strategies summarized in Ref. \cite{tqpt}. 

The presence of edge states in the bulk gap above half-filling Ref. \cite{edge-Bragg, edge-Sarma, edge-Nathan} is an intriguing concept. For this undoubtedly challenging endeavor one would need to engineer a sharp boundary, with a characteristic length on the order of the lattice constant, in order to stabilize the presence of topological edge-states within the center of the trap~\cite{edge-Nathan}.  Loading the atoms into the edge states can be achieved with external light pulses \cite{edge-Sarma, edge-Nathan}. In addition, the dynamical structure factor $S({\bf q},\omega)$ from light Bragg scattering can also provide a direct way to observe the edge states and bulk states as demonstrated in Ref. \cite{edge-Bragg}. The lack of edge-states in the bulk gaps around half-filling could be an important indicator of the existence of a flat band.

To detect the Klein tunneling the most natural approach seems to be designing a potential barrier for the atoms by optical or magnetic means, and prepare the atoms with a well defined momentum, then see if particles have tunneled through the barrier to the other region. However, a direct confirmation of the Klein multirefringent tunneling would require a launch of a multicomponent mass current. To launch such a mass current, several schemes can be used. For example, one can connect the optical lattice to two reservoirs with different chemical potentials as in \cite{atomtronics1, atomtronics2}, exert a static force from a tilted optical lattice \cite{static-forcing}, or by an effective electric field on the atoms from the optical dipole force \cite{electric-field}. Measuring the mass current across the barrier will consequently reveal the intricate dependence on helicity for the tunneling dynamics.


\section{Conclusions and Outlook}
\label{Conclusions}

In this paper, we have proposed the quantum simulation of Dirac-Weyl fermions with arbitrary spin by a particular laser-assisted tunneling in optical lattices. By tuning the spin-dependent hopping according to the $\mathfrak{su}$(2)
Lie algebra,  we can assign an arbitrary spin $s$ to these fermions, and  go beyond the standard  spin-$\half$ regime of high-energy physics. 

We have presented a detailed study, both analytically and numerically,  of  several striking aspects of Dirac-Weyl fermions. In particular, our system hosts two different phases: a semi-metallic phase for half-integer $s$, and a metallic phase that contains a flat zero-energy band for integer spin $s$. 
We have shown that the low-energy transport in the semi-metallic phase is characterized by multirefringent   spin 1/2 Dirac-Weyl fermions moving at different speeds. As a consequence, we also find an exotic Klein tunneling  across 
a potential barrier.  In the presence of a synthetic magnetic field, we have connected the
Weyl-Landau problem to the Dicke model known from quantum optics. The corresponding Hamiltonian presents a rich structure of Weyl-Landau levels 
 and zero-energy modes whose robustness can be related to an index theorem for half-integer $s$, which also includes an anomalous half-integer quantum Hall effect.

Many interesting avenues remain however unexplored. It is possible to  
engineer a mass term by a particular on-site Raman transition~\cite{dirac_fermions_2_3} in order to explore a massive regime and even look for a topological insulating phase. Another interesting possibility is to engineer a curved spacetime background following~\cite{dirac_curved}, and study the effects of the higher spin of these Dirac-Weyl fermions. One can also separate each  spin 1/2 Dirac-Weyl component of the Dirac-Weyl fermions with high spin  by an appropriate tailoring of the spin-dependent hopping \cite{tqpt,shen:2010}, which could potentially lead to interesting topological quantum phase transitions. An intriguing supersymmetric algebraic structure~\cite{klein_supersymmetry} also deserves further attention in the  Klein multi-refringent regime. We thus believe that several new effects can be explored starting from the results presented in this work.

{\bf Acknowledgements.} Z. L acknowledges 
the financial support from SUPA (Scottish Universities Physics Alliance). A.B. thanks  MICINN  FIS2009-10061, 
CAM QUITEMAD, European FET-7 PICC and HIP, UCM-BS GICC-910758, and acknowledges very useful discussions with L. Mazza.
N.G. thanks the F.R.S-F.N.R.S. for financial support, and also V. Wens, F. Gerbier, D. Bercioux and J. Dalibard for interesting discussions. P.\"O acknowledges support from the Carnegie Trust for the Universities of Scotland. We thank R. Shen for an interesting discussion regarding the boundary conditions for the Klein tunneling. 

 {\it Note added.} After the submission of this work, another manuscript appeared which explored the same idea of Dirac-Weyl fermions with high spin but in a different setup  \cite{Dora}.

\appendix

\section{Experimental realization of spin-dependent hopping}
\label{appendix_exp}

\label{appendix_exp}
\begin{figure}[!hbp]
	\centering
	\includegraphics[width=0.7\columnwidth]{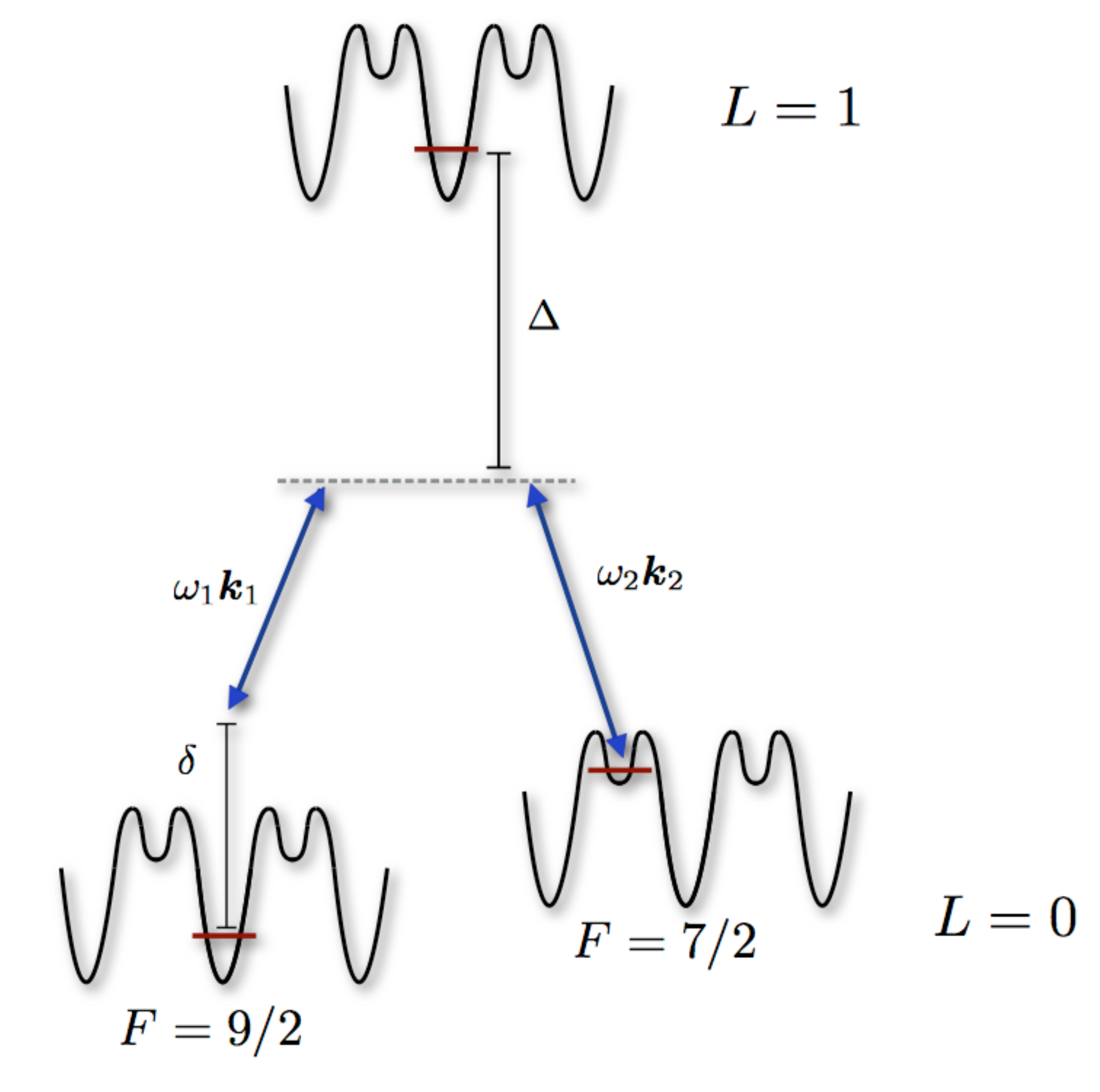}
	\caption{{\it Superlattice scheme for a laser-assisted hopping}. Atoms in the ground state $L=0, F=9/2$ are coupled via a two-photon Raman transition to neighboring atoms in a different hyperfine manifold $L=0, F=7/2$. This Raman transition takes place through an intermediate excited state $L=1$, such that for large enough detuning $\delta,|\Omega_1|,|\Omega_2|\ll \Delta$, where $\Omega_i$ are the Rabi frequencies for each laser,  one readily arrives at the Hamiltonian in Eq.~\eqref{raman}.}
	\label{superlattice_raman}
\end{figure}

In this Appendix, we describe the main ingredients of a method to engineer the spin-dependent tunneling~\cite{superlattice_scheme} that leads to the Hamiltonian in Eq~\eqref{hamiltonian}. We consider a cloud of ultracold $^{40}$K atoms described by the fermionic field operators $c_{\boldsymbol{x}\tau}^{\dagger}(c_{\boldsymbol{x}\tau})$, where $\tau\in\{1...N\}$ is the internal index that labels a particular subset of Zeeman sublevels in the ground state $L=0,F=9/2$, and $\boldsymbol{x}$ stands for the sites of a two-dimensional square superlattice.  This particular superlattice  follows from the optical potential created by two pairs of counter-propagating lasers along each axis $\alpha$, $V({x}_{\alpha})=V_1\sum_{j}\cos^2(k_{\text{L}}x_{\alpha})+V_2\sum_{j}\cos^2(2k_{\text{L}}x_{\alpha})$, where $V_1\gg V_2$ represent the lattice depths, and $k_{\text{L}}$ an optical wavevector. As shown in Fig. \ref{superlattice_raman}, the atoms are trapped in a periodic structure of  {\it primary} and {\it secondary} minima, that shall be referred to as {\it sites} $\boldsymbol{r}$, and {\it links} $\boldsymbol{l}$ henceforth. For a deep optical superlattice, the atomic tunneling between neighboring sites will be completely suppressed. The fundamental idea to engineer a spin-dependent hopping is to assist this tunneling using additional lasers that drive a Raman transition to an excited state in the hyperfine manifold $F=M=7/2$ (see Fig.~\ref{superlattice_raman}), here represented by the fermionic operators $d_{\boldsymbol{x}}^{\dagger}(d_{\boldsymbol{x}})$.  The use of a pair of lasers in a Raman configuration is two-fold. On the one hand, the effective frequency can be  tuned to the microwave transition $F=9/2\leftrightarrow F=7/2$. On the other hand, the effective wavevector can be large so that the lasers impart enough momentum to the atoms to tunnel between neighboring lattice sites. As customary~\cite{cohen_book}, such a two-photon transition is obtained after the adiabatic elimination of a higher excited state in the fine structure $L=1$.
The Raman lasers aligned along a particular hopping direction, not only drive the transition between the hyperfine levels, but also transfer a finite momentum to the atoms which allows them to tunnel to a neighboring site. This assisted hopping~\cite{synthetic_gauge} is described by the following Hamiltonian
\begin{equation}
\label{raman}
H_{\text{R}}=\sum_{\boldsymbol{x}',\boldsymbol{x}}\Omega^{\text{eff}}_{\tau}S_{\boldsymbol{x}\boldsymbol{x}'}c_{\boldsymbol{x}'\tau}^{\dagger}d_{\boldsymbol{x}}\ee^{-\ii\omega_{\text{R}}t}+\text{H.c},
\end{equation}
where $\Omega^{\text{eff}}_{\tau}$ is the two-photon Rabi frequency driving the transition $\ket{F=9/2,\tau}\leftrightarrow\ket{F=7/2,7/2}$, and $S_{\boldsymbol{x}\boldsymbol{x}'}=\bra{\boldsymbol{x}}\ee^{\ii\boldsymbol{k}_{\text{R}}\boldsymbol{r}}\ket{\boldsymbol{x}'}=\int d^3r w^*_{\boldsymbol{x}}(\boldsymbol{r})\ee^{\ii\boldsymbol{k}_{\text{R}}\boldsymbol{r}}w_{\boldsymbol{x}'}(\boldsymbol{r})$ determines the momentum transfer that the Raman lasers impart on the atom, thus assisting the transition between neighboring superlattice sites. The parameters of this two-photon Raman transition  $\omega_{\text{R}}=\omega_1-\omega_2$, and $\boldsymbol{k}_{\text{R}}=\boldsymbol{k}_1-\boldsymbol{k}_2$, follow from each laser frequency and wavevector. Here, we have introduced the corresponding Wannier functions $w_{\boldsymbol{x}}(\boldsymbol{r})$. In the expression of $S_{\boldsymbol{x}\boldsymbol{x}'}$, one sees the importance of using a two-photon Raman scheme rather than a simple microwave, since the integral  between neighboring Wannier functions will only be finite when the imparted momentum is large (i.e. the effective wavelength is on the order of the lattice spacing $\lambda_{\rm R}\approx a$, typically a few hundred nanometers).  As shown in~\cite{superlattice_scheme}, by selecting an appropriate detuning, Zeeman splitting, and lattice staggering, it is possible to perform an adiabatic elimination of the auxiliary states $d_{\boldsymbol{x}}^{\dagger}(d_{\boldsymbol{x}})$ that reside on the links, and thus obtain an effective Hamiltonian that describes the hopping of $F=9/2$ atoms along the primary sites of a two-dimensional lattice.  Therefore, the auxiliary link serves as a bus that allows us to assist the tunneling, and one obtains the effective  Hamiltonian in Eq.~\eqref{hamiltonian}
\begin{equation}
H=-\sum_{\boldsymbol{r},\boldsymbol{\nu}}\sum_{\tau\tau'}t_{\nu}[\mathbb{T}_{\nu}]_{\tau'\tau}c_{\boldsymbol{r}+{\boldsymbol{\nu}},\tau'}^{\dagger}c_{\boldsymbol{r}\tau}+\text{H.c.},                           
\end{equation} 
where the tunneling strengths $t_{\nu}$ now depend on the four-photon Rabi frequencies.  We note that the particular matrices $\mathbb{T}_{\nu}$ can in principle be designed at will, although the experimental requirements are certainly challenging. 

 For the simplest situation $s=1/2$, one  selects a pair of Zeeman sub-levels $M=9/2,7/2$, and the hopping operators would correspond to $\mathbb{T}_x=\half\tau_x,\mathbb{T}_y=\half\tau_y$. Each of these can be engineered with a single pair of Raman beams, once their frequencies are tuned to the resonance given by the Zeeman splitting and the lattice staggering. To account for the real/complex matrix elements, one should control the laser phases appropriately, which can be accomplished by means of acusto-optical modulators. The scheme gets more complicated for  $s=1$, where the hopping operators have now four non-vanishing elements, and thus double the number of beams required. Let us note, however, that since the different resonance frequencies rely on the Zeeman splitting and staggering, which is a small fraction of the laser frequency~\cite{superlattice_scheme}, the desired frequencies can be obtained from the same source, once the beams are split, and their frequencies  tuned by an acusto-optical modulator. For arbitrary $s$, the scheme is more involved, yet benefits from the fact that the hopping matrices are of sparse nature. Let us finally remark that there might be more clever schemes that take adventage of the light polarization to select the different hopping elements.

\section{Dirac-Weyl fermions and topological invariants}
\label{appendix_weyl}

In this Appendix, we show how to describe a single Dirac-Weyl fermion with high spin as a collection of   spin 1/2 Dirac-Weyl fermions, and also give an explicit derivation of the topological charges that can be assigned to these excitations. The Hamiltonian in Eq.~\eqref{band_model} can be diagonalized using a similarity transformation for the  $\mathfrak{su}(2)$ Lie algebra~\cite{puri_book}. In particular,
the $\mathfrak{su}(2)$ rotation  
\begin{equation}
U=\ee^{\ii\frac{\pi}{2}(n_{\boldsymbol{k}}^yS_x-n_{\boldsymbol{k}}^xS_y)}, \hspace{2 ex} n_{\boldsymbol{k}}^{\nu}=-2t_{\nu}\cos k_{\nu}/\epsilon(\boldsymbol{k}),
\end{equation}
brings the Hamiltonian into a diagonal form 
\begin{equation}
\label{diag}
H_{\boldsymbol{k}}'=U^{\dagger}H_{\boldsymbol{k}}U=E_0P_0+\sum_{m>0}E_m(\boldsymbol{k})\tilde{\sigma}^z_m,
\end{equation}
 where the energies $E_m(\boldsymbol{k})$ are those of Eq.~\eqref{bands}, and we have defined $P_0=\ket{s,0}\bra{s,0}$ as the projector onto the zero-energy flat-band for $m=0$, and the Pauli matrices  $\tilde{\sigma}^z_m=\ket{s,m}\bra{s,m}-\ket{s,-m}\bra{s,-m}$, where $\ket{s,m}$ are the eigenstates of $S_z$. Let us note that the eigenstates of the Hamiltonian $\ket{E_m}=U\ket{s,m}$ can be understood as the $\mathfrak{su}(2)$ spin-coherent states for different fiducial states.

 Once we have an analytic expression for the diagonal Hamiltonian, we shall show that for half-integer spin, each Dirac-Weyl fermion with high spin can be expressed as a collection of  spin 1/2 Dirac-Weyl fermions. Let us define the Pauli matrices  $\tilde{\sigma}^x_m=\ket{s,m}\bra{s,-m}+\text{H.c.}$, and $\tilde{\sigma}^y_m=-\ii\ket{s,m}\bra{s,-m}+\text{H.c.}$. Now, we perform the following block-diagonal rotation to the Pauli matrices  in Eq.~\eqref{diag}
 \begin{equation}
 U'=\text{diag}(\ee^{-\ii\frac{\pi}{2}\epsilon(\boldsymbol{k})(-n_{\boldsymbol{k}}^y\tilde{\sigma}^x_1+n_{\boldsymbol{k}}^x\tilde{\sigma}^y_1)},...,\ee^{-\ii\frac{\pi}{2}\epsilon(\boldsymbol{k})(-n_{\boldsymbol{k}}^y\tilde{\sigma}^x_N+n_{\boldsymbol{k}}^x\tilde{\sigma}^y_N)    }).
 \end{equation}
 The block structure of the Hamiltonian~\eqref{diag}, allow us to easily derive the effective description in terms of  spin 1/2 Dirac-Weyl fermions
 \begin{equation}
 \label{particle_hole_ham}
 H''_{\boldsymbol{k}}=U'H'_{\boldsymbol{k}}U'^{\dagger}=-\sum_{m>0}(2t_xm\cos k_x\tilde{\sigma}^x_m+2t_ym\cos k_y\tilde{\sigma}^y_m),
 \end{equation}
 which directly leads to the  Weyl-like Hamiltonian in Eq.~\eqref{dirac_ham}. We note that the Dirac-Weyl spinor introduced here is built from pairs of states with opposite spin $\tilde{\Psi}_m(\boldsymbol{p})=U'U(\ket{s,m},\ket{s,-m})^t$. By a simple counting argument, one finds that there are $N_{\text{D}}=s+\half$ underlying   spin 1/2 Dirac-Weyl femions per Dirac-Weyl fermion of half-integer spin. Conversely, for the integer-spin case, one finds $N_{\text{D}}=s$  spin 1/2 Dirac-Weyl fermions and a single zero-energy flat band. \\
 
 We now derive the Berry phases~\cite{berry_phase} that can be assigned to the semimetallic phase. The parallel transport of  the  eigenstates   $\{\ket{E_m(\boldsymbol{k})}\}$ below Eq.~\eqref{diag} can be defined from the so-called Berry connection on the principal fiber bundle $\mathcal{P}(\mathbb{T}_2,U(1))$, where $\mathbb{T}_2$ is the 2-torus assigned to the Brillouin zone. The Berry connection is defined as $\boldsymbol{\mathcal{A}}_m(\boldsymbol{k})=\ii\bra{E_m(\boldsymbol{k})}\nabla_{\boldsymbol{k}}\ket{E_m(\boldsymbol{k})}=\ii\bra{s,m}U^{\dagger}_{\boldsymbol{k}}\nabla_{\boldsymbol{k}}U_{\boldsymbol{k}}\ket{s,m}$. This expression can be evaluated using Sneddon's formula for the parametric differentiation of the operator exponential~\cite{puri_book}. In the particular case of eigenstates at half-filling, the Berry connection reads  $\boldsymbol{\mathcal{A}}_{-|m|}(\boldsymbol{k})=|m|(n_{\boldsymbol{k}}^x\nabla_{\boldsymbol{k}}n_{\boldsymbol{k}}^y-n_{\boldsymbol{k}}^y\nabla_{\boldsymbol{k}}n_{\boldsymbol{k}}^x)$. We can now derive the Berry phase 
$\gamma_{m}=\oint_C \text{d}\boldsymbol{k}\cdot\boldsymbol{\mathcal{A}}_m(\boldsymbol{k})$, where $C$ represents a cycle in the Brillouin zone that surrounds a certain Dirac  singularity.  Performing the corresponding line integral, we have found that the Berry phase  associated to each filled eigenstate is $\gamma^{\boldsymbol{d}}_{-|m|}=-2\pi|m|\text{sgn}(d_xd_y)$, where $d_x,d_y\in\{-1,1\}$ determine the particular Dirac  point. Since the Berry phase is defined modulo $2\pi$, we find that the Berry phase for each of the filled bands has an abnormal parity $\gamma^{\boldsymbol{d}}_{-|m|}=\pi$ for half-integer spin $s$.  This particular homotopic and topological invariant underlies many of the fascinating phenomena of graphene~\cite{graphene_rev}.

Finally, we close this appendix by calculating an integer-value topological charge which is more descriptive than the Berry phase presented above. This topological invariant~\cite{top_charge}, is defined for particle-hole symmetric Hamiltonians, such as Eq.~\eqref{particle_hole_ham}, which fulfills $\{H''_{\boldsymbol{k}},\Gamma\}=0$, where $\Gamma=\text{diag}\{\tilde{\sigma}^z_1,...\tilde{\sigma}^z_N\}$ exploits the block structure of the Hamiltonian and fulfills $\Gamma^{\dagger}=\Gamma,\Gamma^2=\mathbb{I}$. The topological invariant is defined as a winding number $\mu=\frac{1}{4\pi i}\oint_C\text{Tr}\{\Gamma (H_{\boldsymbol{k}}'')^{-1}\text{d}H''_{\boldsymbol{k}}\}$, which vanishes unless the cycle $C$ encloses a zero of the Hamiltonian. Therefore, this invariant is ideally suited to detect the Dirac  points in the Fermi surface, and the value it assigns $\mu_{\boldsymbol{d}}\in\mathbb{Z}$ shall be topologically protected for any small perturbation that preserves the particle-hole symmetry. This winding number reduces to $\mu^{\boldsymbol{d}}=\sum_{m>0}\mu^{\boldsymbol{d}}_{-|m|}$, where 
\begin{equation}
\mu^{\boldsymbol{d}}_{m}=\frac{1}{4\pi i}\oint_C(\text{d}\log h_{m\boldsymbol{k}}-\text{d}\log h^*_{m\boldsymbol{k}}), 
\end{equation}
and we have introduced $h_{m\boldsymbol{k}}=2t_xm\cos k_x+2\ii t_ym\cos k_y$. Performing the line integrals, it can be shown that the topological charges around the Dirac  points read $\mu^{\boldsymbol{d}}=N_{\text{D}}\text{sgn}(d_xd_y)$, where $N_{\text{D}}=(2s+1)/2$. Accordingly, the charges are integer values, and their sign depends on the particular Dirac  point $d_x,d_y\in\{-1,1\}$ under consideration. For particle-hole preserving perturbations, only when two opposite charges meet, a quantum phase transition  can take place. Therefore, on very general grounds, we can claim that the semimetallic phase is topologically protected.

\section{Analytical solution of the Weyl Landau levels, topological zero-energy modes and the Index theorem }
\label{appendix_LL}

In this Appendix, we derive the exact solution of the Hamiltonian in Eq.~\eqref{dicke_model}, and describe the appearance of multiple zero-energy modes which are responsible for the half-integer anomaly in the quantum Hall sequence.

{\it Complete analytical solution: } In this part of the Appendix, we derive the exact solution of the Hamiltonian in Eq.~\eqref{dicke_model} for the Dirac  point  $\boldsymbol{d}=(+1,+1)$, and  in the isotropic regime $c_x=c_y$. To do this, we write out the matrix elements explicitly as  $H^{\boldsymbol{d}}_{\text{WL}}=g_{\mathbf{d}+}\sum_{n,m}[E_{nm}\delta_{n,m+1}\rho_ma^{\dag}+E_{nm}\delta_{n,m-1}\rho_na]$,
where $\rho_n=\sqrt{n(2s+1-n)}$  and $E_{nm}$ is the $(2s+1)\times(2s+1)$ matrix with 1 at row $n$ and column $m$, and zeros otherwise. The 
eigenvector can be expressed as $\Psi=(\phi_1,\phi_2,\cdots,\phi_{2s+1} )^t$, and the eigen-equations are the following 
operator equations $\rho_{i-1}a^{\dag}\phi_{i-1}-\epsilon\phi_i+\rho_ia\phi_{i+1}=0$ with $i=1,2,\cdots,2s+1$. We solve the Weyl-Landau levels by 
successive substitutions, i.e., we substitute the second equation into the first one, and then substitute the 
third one to the result obtained by the previous substitution and repeat the process (see Table \ref{substitutions}). After $r$ 
substitutions, we get $A_{r,i}(aa^{\dag})^{i-1}\phi_r=\rho_rB_{r,j}(aa^{\dag})^{j-1}a\phi_{r+1}$, where $A_{r,i}$ and $B_{r,j}$ are energy-dependent 
constants, using the Einstein summation convention for repeated indexes $i$,$j$. Setting
$\rho_ra^{\dag}\phi_r=\epsilon\phi_{r+1}-\rho_{r+1}a\phi_{r+2}$ , we get $A_{r,i}(a^{\dag}a)^{i-1}(\epsilon\phi_{r+1}-\rho_{r+1}a\phi_{r+2})=B_{r,j}\rho_r(a^{\dag}a)^j\phi_{r+1}$. 
Recasting the expression, we obtain
\begin{widetext}
\begin{equation}
\begin{split}
A_{r+1,i}&=\epsilon\sum_{[(r+2)/2]\geq k\geq i}A_{r,k} {k-1\choose i-1}(-1)^{k-i}-\rho_r^2\sum_{[(r+1)/2]\geq k\geq i-1}B_{r,k} {k\choose i-1}(-1)^{k-i+1}, \hspace{1ex} B_{r+1,i}=\sum_{[(r+2)/2]\geq k\geq i}A_{r,k} {k-1\choose i-1}(-1)^{k-i},
\end{split}
\end{equation}
\end{widetext}
where ${n\choose m}$ is a binomial coefficient and $[r]$ is the integer part of $r$. By using the cutoff condition $\rho_{2s+1}=0$ , the Weyl-Landau levels are obtained compactly as  
\begin{eqnarray}
\sum_{i=1}^{[s+3/2]}A_{2s+1,i} (n+1)^{i-1}=0
\end{eqnarray} 
where $A_{2s+1,i}$ are determined by the recursion above with initial values $A_{1,1}=\epsilon$ and  $B_{1,1}=1$. Here, 
n is the eigenvalue of the number operator $a^{\dag}a$ for the Fock state $\ket{n}$.
The eigenvector is then expressed as:  
$ \Psi_n^s=(f_{2s}(n,\epsilon)\ket{n-2s},f_{2s-1}(n,\epsilon)\ket{n-2s+1},\cdots,f_0(n,\epsilon)\ket{n})^t $, where 
 $f_i(n,\epsilon) $ are determined by the algebraic equation 
 \begin{equation}
 \rho_{i-1}f_{2s+2-i}\sqrt{n-2s+i-1}-\epsilon f_{2s+1-i}+\rho_if_{2s-i}\sqrt{n-2s+i}=0,
 \end{equation}
 with  $i=1,2,\ldots,2s+1$. For $s=1/2$ and $s=1$, the results in Table \ref{substitutions} confirm the results in~\cite{klein_tunneling,shen:2010}. However, the general expression here is applicable to any arbitrary spin. The Weyl-Landau 
states of these high spin particles are a mixture of successive 
non-relativistic Landau Levels, i.e. the 
spin and orbital degrees of freedom of $\Psi_n^s$
are highly entangled. This entanglement is a source of many 
interesting topics, like the mesoscopic superposition states in relativistic Landau levels~\cite{dirac_cats}, possible 
fractional Hall states with high Landau-level index and non-fractional to fractional QHE transition~\cite{fqhe_graphene}, and non- 
abelian anyons~\cite{non_abelain_anyons_atoms}. We note that our Weyl-Landau problem can be mapped to a single-mode Dicke 
model which shows a quantum phase transition from the normal to superradiant state (i.e. the ground 
state has a  large Weyl-Landau index n) in the limit of $s\to\infty$. This may be the way forward to realize a stable fractional QHE state with higher Weyl-Landau index $n$ in this kind of system when the interaction is turned on.

\begin{table*}
\caption{\label{substitutions} {\it Weyl-Landau levels by successive substitutions}}
		\begin{tabular}{cccc}
		\hline \hline
		$n^{th}$ substitution & expression & Weyl-Landau levels & pseudospin\\
		\hline \hline
		$1^{st}$ & $(A_{2,1}+A_{2,2}aa^{\dag})\phi_2=\rho_2(B_{2,1})a\phi_3$ \footnote{$A_{2,1}={\epsilon}^2+\rho_1^2,A_{2,2}=-\rho_1^2;B_{2,1}=\epsilon $} & $\epsilon=\pm g\sqrt{2n}$ & $s=1/2$\\
		\hline

$2^{nd}$ & $(A_{3,1}+A_{3,2}aa^{\dag})\phi_3=\rho_3(B_{3,1}+B_{3,2}aa^{\dag})a\phi_4$ \footnote{$         A_{3,1}=\epsilon({\epsilon}^2+2\rho_1^2+\rho_2^2),A_{3,2}=-\epsilon({\rho_1}^2+\rho_2^2);B_{3,1}={\epsilon}^2+2\rho_1^2,B_{3,2}=-\rho_1^2 $} 
& $\epsilon=0,\pm g\sqrt{2(2n-1)}$ & $s=1$\\
		\hline
		$3^{rd}$ & $\left(A_{4,1}+A_{4,2}aa^{\dag}+A_{4,3}(aa^{\dag})^2\right)\phi_4=\rho_4(B_{4,1}+B_{4,2}aa^{\dag})a\phi_5$ \footnote{$A_{4,1}={\epsilon}^4+{\epsilon}^2(3\rho_1^2+2\rho_2^2+\rho_3^2)+3\rho_1^2\rho_3^2,A_{4,2}=-{\epsilon}^2(\rho_1^2+\rho_2^2+\rho_3^2)-4\rho_1^2\rho_3^2,A_{4,3}=-\rho_1^2;B_{4,1}=\epsilon({\epsilon}^2+3\rho_1^2+2\rho_2^2),B_{4,2}=-\epsilon(\rho_1^2+\rho_2^2)$}& $\epsilon=\pm g\sqrt{(5n-1)\pm\sqrt{16(n-1)^2+9}}$ & $s=3/2$\\
		\hline
		$4^{th}$ & $\left(A_{5,1}+A_{5,2}aa^{\dag}+A_{5,3}(aa^{\dag})^2\right)\phi_5=\rho_5\left(B_{5,1}+B_{5,2}aa^{\dag}+B_{5,3}(aa^{\dag})^2\right)a\phi_6$ \footnote{$A_{5,1}=\epsilon({\epsilon}^4+{\epsilon}^2(4\rho_1^2+3\rho_2^2+2\rho_3^2+\rho_4^2)+8\rho_1^2\rho_3^2+4\rho_1^2\rho_4^2+3\rho_2^2\rho_4^2),
           A_{5,2}=\epsilon({\epsilon}^2(\rho_1^2+\rho_2^2+\rho_3^2+\rho_4^2)+6\rho_1^2\rho_3^2+5\rho_1^2\rho_4^2+4\rho_2^2\rho_4^2),A_{5,3}=\epsilon(\rho_1^2\rho_3^2+\rho_1^2\rho_4^2+\rho_2^2\rho_4^2);
B_{5,1}={\epsilon}^2({\epsilon}^2+4\rho_1^2+3\rho_2^2+2\rho_3^2)+8\rho_1^2\rho_3^2,B_{5,2}=-{\epsilon}^2(\rho_1^2+\rho_2^2+\rho_3^2)-6\rho_1^2\rho_3^2),B_{5,3}=\rho_1^2\rho_3^2$}& $\epsilon=0,\pm g\sqrt{5(2n-3)\pm 3\sqrt{4n^2-12n+17}}$ & $s=2$\\
		\hline
		\multicolumn{4}{c}{\vdots}\\
		\hline
		$r^{th}$ & $\sum_i A_{r,i}(aa^{\dag})^{i-1}\phi_r=\sum_j \rho_r B_{r,j}(aa^{\dag})^{j-1}a\phi_{r+1}$  & & \\
		\hline \hline
		\end{tabular}
\end{table*}

{\it Existence of zero-energy modes:} In this part of the Appendix, we explicitly derive the zero-energy modes displayed by the Weyl-Landau
Hamiltonian in Eq.~\eqref{dicke_model}  for $\boldsymbol{d}=(+1,+1)$ in the isotropic regime $c_x=c_y$. The complete Hilbert space is $\mathcal{H}=\mathcal{F}\otimes\mathbb{C}^N$, where $\mathcal{F}=\text{span}\{\ket{n},n=0,1...\}$ is the Fock space of the cyclotron modes, whereas $\mathbb{C}^N=\text{span}\{\ket{s,m},m=-s...,s\}$ is the angular momentum space. Nonetheless, the action of the Hamiltonian decomposes in a set of invariant subspaces $\mathcal{H}=\oplus_{n=0}^{\infty}\mathcal{H}_n$, where each $\mathcal{H}_n$ is spanned by a different combination of the Fock and spin states. In particular, we find that for $n<2s$, there are certain special subspaces with odd dimensionality:
\begin{equation}
\begin{split}
\mathcal{H}_0&=\text{span}\{\ket{0}\ket{s,-s}\},\\
\mathcal{H}_2&=\text{span}\{\ket{2}\ket{s,-s},\cdots,\ket{0}\ket{s,-s+2}\},\\
&\hspace{1ex}\vdots\\
\mathcal{H}_{2s-1}&=\text{span}\{\ket{2s-1}\ket{s,-s},\cdots,\ket{0}\ket{s,s-1}\}.\\
\end{split}
\end{equation}

Each of these subspaces hosts a zero-energy mode of the Weyl-Landau Hamiltonian. By introducing the following quantities $f_{n,s,m}=g\sqrt{n(s(s+1)-m(m+1))}$, where $g$ is the coupling constant introduced in the Weyl-Landau Hamiltonian of Eq.~\ref{dicke_model}, one finds the following zero-energy modes for which $H\ket{E_{0}}=0$: 
\begin{widetext}
\begin{equation}
\label{zero_spinor}
\begin{split}
\ket{E_{0}^{(1)}}&=\ket{0}\ket{s,-s},\\
\ket{E_{0}^{(2)}}&\propto f_{1,s,-s+1}\ket{2}\ket{s,-s}-f_{2,s,-s}\ket{0}\ket{s,-s+2},\\
&\hspace{1ex}\vdots\\
\ket{E_{0}^{(N_{\text{z}})}}&\propto\prod_{n'\text{odd}}^{2s-2} f_{n',s,s-1-n'}\ket{2s-1}\ket{s,-s}-f_{2s-1,s,-s}\prod^{2s-4}_{n'\text{odd}}f_{n',s,s-1-n'}\ket{2s-3}\ket{s,-s+2}+\cdots+\\
&+(-1)^{s-3/2}f_{1,s,s-2}\prod_{n'\text{even}}^{2s-3}f_{2s-1-n',s,-s+n'}\ket{2}\ket{s,s-3}+(-1)^{s-1/2}\prod_{n'\text{even}}^{2s-1}f_{2s-1-n',s,-s+n'}\ket{0}\ket{s,s-1},
\end{split}
\end{equation}
\end{widetext}
where we have omitted an irrelevant normalization factor. By simple counting, we find that the Weyl-Landau Hamiltonian hosts $N_{\text{z}}=s+\half$ zero-energy modes for half-integer spin. Interestingly, the number of zero-modes coincides with the number of the spin 1/2 Dirac-Weyl fermions contained by a single high spin Dirac-Weyl fermion. Therefore, one may argue that each underlying  spin 1/2 Dirac-Weyl fermion hosts a single zero-energy mode, which shall be responsible for a half-integer anomaly in the quantum Hall response of the system (see Eq.~\ref{qhe}). Conversely, the number of zero modes for integer spin is not bounded (see Table~\ref{WLL_spectrum}).

{\it Index theorem and robustness of zero modes:} We have already seen two physical manifestations of topology, namely, the topological charge, or the abnormal Berry phase, that can be assigned to each of the Dirac-Weyl fermions (Sec.~\ref{spectrum}), and the Chern numbers that determine the quantum Hall response of the system (Sec.~\ref{WLLs and AQHE}).  In this part of the Appendix, we describe yet another manifestation of topology: the relation of the zero modes  to the  Atiyah-Singer theorem~\cite{atiyah_index}. This famous theorem, which relates the analytical and topological features of differential operators, has important consequences on the properties of Dirac-Weyl fermions subjected to external gauge fields~\cite{qhe_graphene}. Graphene therefore has turned out to be an excellent platform to understand this relationship both from a theoretical~\cite{pachos_index} and experimental viewpoint~\cite{qhe_graphene}. In this Appendix, we describe how these concepts can be generalized to Dirac-Weyl fermions of arbitrary spin $s$, and we find that only the zero modes of half-integer spin Dirac-Weyl fermions are protected by the topological features of the system. As discussed in Sec.~\ref{WLLs and AQHE}, this justifies the absence of the half-integer anomaly for integer-spin Dirac-Weyl fermions.

The Weyl-Landau Hamiltonian in Eq.~\eqref{dicke_model} for $\boldsymbol{d}=(+1,+1)$ in the isotropic regime $c_x=c_y$ presents the following particle-hole symmetry 
\begin{equation}
 \Gamma_s=\ee^{\ii \pi(S_z+s)}, \hspace{2ex} \{H_{\text{WL}}^{\boldsymbol{d}},\Gamma_s\}=0,
\end{equation}
which fulfills $\Gamma_s^2=\mathbb{I},\Gamma_s^{\dagger}=\Gamma_s$. This operator, known as an involution~\cite{nakahara_book,thaller_book},  allows us to decompose the Hilbert space as $\mathcal{H}=\mathcal{H}_+\oplus\mathcal{H}_-$, where $\mathcal{H}_{\pm}$ follow from the orthogonal projections $P_{\pm}=\half(\mathbb{I}\pm \Gamma_s)$ associated to the $\pm 1$ eigenvalues of the involution. With this formulation, the Weyl-Landau Hamiltonian can be rewritten as  a supercharge $\mathcal{Q}$
\begin{equation}
H_{\text{WL}}^{\boldsymbol{d}}=\mathcal{Q}=\left(\begin{array}{cc}0 & D^{\dagger} \\ D & 0\end{array}\right),
\end{equation}
where the differential operators $D=P_-H_{\text{WL}}^{\boldsymbol{d}}P_+:\mathcal{H}_+\to\mathcal{H}_-$, and  $D^{\dagger}=P_+H_{\text{WL}}^{\boldsymbol{d}}P_-:\mathcal{H}_-\to\mathcal{H}_+$, join the orthogonal subspaces. In this language, the {\it analytical index} of the supercharge can be expressed as 
\begin{equation}
\text{ind}\mathcal{Q}=\nu_+-\nu_-=\text{dim}(\text{ker}D)-\text{dim}(\text{ker}D^{\dagger}).
\end{equation}
 For elliptic operators~\cite{nakahara_book}, this index can be related to the {\it topological features} of the system via the famous Atiyah-Singer index theorem. This relationship not only gives insight into the number of zero modes in the system, but also pinpoints their robustness with respect to local perturbations of the Hamiltonian.  From the results in Tables~\ref{WLL_spectrum} and~\ref{WLL_zero_modes}, we observe that 
\begin{equation}
\hspace{1ex} \nu_+=s+\half,\hspace{1ex} \nu_-=0, \hspace{2ex} s \text{  half-integer},
\end{equation}
and therefore the total number of zero modes determines the index $\text{ind}\mathcal{Q}=s+\half$, which is related to the total magnetic flux that pierces the system. Therefore, these zero modes are extremely robust with respect to local perturbations of the Hamiltonian. Conversely, the number of zero modes for integer spin is unbounded (see Table~\ref{WLL_spectrum}). In this case, the differential operator $D$ is not an elliptical operator~\cite{comment2}, and thus the Index theorem does not apply. Accordingly, the zero modes for an integer-spin Dirac-Weyl fermion are not topologically protected.

\section{Flat band, zero modes and Hall conductivity}

\label{appendixnath}

In Section \ref{WLLs and AQHE}, we have demonstrated that Dirac-Weyl particles with integer spin $s$ do not present the anomalous (half-integer) quantum Hall effect. Namely, we have shown the absence of edge-states stemming from the zero modes, therefore leading to a zero Hall plateau above half-filling. This effect is particularly interesting as it seems to be rooted in the absence of an index theorem that guaranties the robustness of these topological zero modes. An alternative explanation comes from the possibility that edge-states with {\it zero} velocity could be hidden in the flat band and would therefore not contribute to the Hall conductivity. 

The squeezing of the edge-states, namely the fact that their dispersion relation is given by $E(k)=0$ for integer $s$, can be investigated in a model that extrapolates continuously between the case of Dirac-Weyl particles with integer and half-integer spin. Such a model has been introduced by Kennett et al. \cite{Kennett2010}. This model has four sites in its unit cell and contains a fundamental parameter $\beta$: When $\beta=1$, the model is equivalent to the Lieb lattice \cite{goldmanlieb}, therefore describing a spin-1 Dirac-Weyl particle and displaying a flat band. For arbitrary $\beta$, the energy spectrum displays two interpenetrating cones and thus describes a spin-$3/2$ Dirac-Weyl particle.

We have computed the Hall conductivity for this model, and we find that $\sigma_H=\pm e^2/h$ around half-filling for $\beta \ne 1$. This is in perfect agreement with the fact that the model displays a single Dirac point, $N_D=1$ with $N_Z=2$ zero modes (cf. Eq. \eqref{qhe}). For $\beta=1$, one observes a zero Hall conductivity plateau at half-filling, as expected for a spin-1 Dirac-Weyl particle. Therefore, this model is well suited to investigate the fate of the edge-states as one goes from $\sigma_H=\pm e^2/h$ (half-integer spin) to $\sigma_H=0$ (integer spin) around half-filling.

Thus, we now investigate the edge-states in this model. We have computed the energy spectrum of Kennett's model in the presence of an external magnetic field, using a  cylindrical geometry. For $\beta=0$, the first bulk gap hosts a single edge-state. As $\beta$ is increased, the energy curves $\tilde{E}(k)$ around $E=\pm 1$ are progressively squeezed while opening an energy gap. For $\beta=1$, this energy gap is exactly located around $E=0$ and the energy curves $\tilde{E}(k)$ disappear into the flat band, i.e. they are completely squeezed  $\tilde{E}(k)=0$. Therefore, the edge-states associated to these curves, hiding at $E=0$, have a zero velocity in the limit $\beta \rightarrow 1$, and thus cannot contribute to the Hall conductivity. 

This analysis emphasizes the existence of edge-states with zero velocity within the flat band, in the presence of a magnetic field. While we have demonstrated this property for Kennett's model, which is equivalent to the Lieb lattice in the limit $\beta \rightarrow 1$, we believe that it should be applicable to general systems exhibiting Dirac-Weyl particles with integer spin.

\section{Klein paradox and multi-refringence}

\label{appendix_klein}

In this Appendix we give a general description of the Klein multi-refringence of the Dirac-Weyl fermions with arbitrary spin. We shall consider the isotropic, $c_x=c_y=c$, Weyl-like Hamiltonian in Eq.~\eqref{weyl_ham}, which is valid for low-momentum excitations around the Dirac  point $\boldsymbol{d}=(+1,+1)$, i.e.,  $H_{\text{W}}^{\boldsymbol{d}}(\boldsymbol{k})=\sum_{\nu}c \hbar  S_{\nu}k_{\nu}$.  For simplicity, we consider a spin-$s$ Dirac-Weyl particle tunneling through a rectangular potential barrier with 
potential $V_0$ in the interval $0<x<D$ and zero elsewhere. The particle is incident on the interface at $x=0$ 
at an angle $\theta$ from the interface normal (see Fig.~\ref{klein1} ). The plane-wave part of the solution is  $\ee^{\ii(k_xx+k_yy)}$, where $k_x=k\cos\theta$ and $k_y=k\sin\theta$. 
Due to the particle-hole symmetry, the helicities form $[s+1/2]$ pairs, where $[s]$ stands for the integer part of s.
For incident particles with energy $V_0>E>0$, and 
helicity $h_0$ (i.e. the helicity determines the corresponding energy $E=c\hbar kh_0)$), all negative-helicity channels are coupled inside the potential barrier with the 
relation $(E-V_0)=c\hbar k_{in}^hh$, and $k_{in,x}^h=k_{in}^h\cos\phi_{in}^h$ and $k_{in,y}^h=k_{in}^h\sin\phi_{in}^h$ . Outside the potential barrier, 
helicity $h_0$ is allowed to convert to other positive values under the condition of energy and momentum 
conservation, i.e., 
  $k_{out}^hh=kh_0$ and $k_{out,x}^h=k_{out}^h\cos\phi_{out}^h$ and $k_{out,y}^h=k_{out}^h\sin\phi_{out}^h$. Due to the conservation 
of parallel wavevectors in the tunneling process, one has $k_{in}^h\sin\phi_{in}^h=k \sin\theta= k_{out}^h\sin\phi_{out}^h$. Note that a non-zero 
helicity is not allowed to convert into a zero helicity due to the violation of energy conservation. The wave 
function in the three regions is: 
\begin{eqnarray}
\Psi_I&=&\phi_R^{h_0}\ee^{\ii k_xx}+\sum_{h=\frac{1}{2}  \text{or} 1}^{h=s}r_h\phi_L^h\ee^{-\ii k_{out,x}^hx},\quad x<0\nonumber \\
\Psi_{II}&=&\sum_{h=-s}^{h=-\frac{1}{2} \text{ or } -1}\!\!\!\!\!A_h\phi_R^h\ee^{\ii k_{in,x}^hx}+B_h\phi_L^h\ee^{-\ii k_{in,x}^hx}, \quad 0<x<D\nonumber\\
\Psi_{III}&=&\sum_{h=\frac{1}{2}\text{or} 1}^{h=s}t_h\phi_R^h\ee^{\ii k_{out,x}^hx}, \quad D<x \nonumber
\end{eqnarray}
where the spinor $\phi_R^h(\phi_L^h)$
is the eigenvector of $H_{\text{W}}^{\boldsymbol{d}}$ with helicity $h$ for the right (left) moving wave. 
Note that $\phi_{in}^h$ and $\phi_{out}^h$
for left moving waves are obtained by using the $(\pi-\phi_{in}^h)$ and $(\pi-\phi_{out}^h)$ angles in the solution for the right moving wave. Here, $r_h,A_h,B_h$ and $t_h$ are $4[s+1/2]$ unknown parameters to be determined. By 
integrating the equation  $H_{\text{W}}^{\boldsymbol{d}}\Psi=E\Psi$ over an interval in the vicinity of the interface, the 
boundary conditions are obtained. For half-integer spin, the boundary conditions require the continuity of 
each component of the ($2s+1$)-component spinor at the two boundaries of $x=0$ and $x=D$, which give $4s+2$  equations that equal to the number of unknowns, $4[s+1/2]=4s+2$ for half-integer spin. 
However,  for integer spin where even and odd components of the spinor are decoupled, the boundary conditions require the continuity of each even spinor component, together with  the continuity of a sum of two neighbouring odd components  at  $x=0$ and $x=D$ (specifically, continuity 
of $\rho_{2i-1}\Psi_{2i-1}+\rho_{2i}\Psi_{2i+1}$ with $i=1,\ldots,s$). These conditions  give $4s$ equations 
that also equal to the number of unknowns, $4[s+1/2]=4s$ for integer spin. Thus, in principle, the 
tunneling properties can be completely determined by the wavefunction and boundary conditions.  
Since different helicity spinors carry a different current, the transmission and reflection coefficients have to be 
renormalized with respect to the incident current, i.e.,$T_h=t_h^2[\phi_R^{h\dag}S_x\phi_R^h]/[\phi_R^{h_0\dag}S_x\phi_R^{h_0}]$  
and $R_h=t_h^2[\phi_L^{h\dag}S_x\phi_L^h]/[\phi_R^{h_0\dag}S_x\phi_R^{h_0}]$
according to the current density $j=\Psi^{\dagger}S_x\Psi$ of the Weyl-like Hamiltonian.
Consequently, conservation of the current requires $\sum_{1/2,1}^{h=s}(R_h+T_h)=1$. In general, for incident particles with spin-$s$ Dirac-Weyl fermions and  fixed helicity, the transmission show $[s+1/2]$-fringence.

{\it Role of evanescent waves:} In graphene, evanescent waves don't play a role in the Klein tunneling due to the single-layered cone structure. For multiple-layered cones however, evanescent waves will play an important role in the coupling and transformation of one helicity to another both inside and outside the barrier, by exerting cutoff conditions for each helicity component. We first consider the coupling of the incident helicity to the ones inside the barrier.  For a low enough barrier, there are negative helicities inside the barrier which are not coupled; they are evanescent. Specifically,  for an incident wave beyond the critical angle $\theta_c=arcsin((V_0-E)/c\hbar k m) $ where $m=1/2$ or $1$, all helicities are uncoupled, thus there is no transmission. For higher barriers, beyond the critical angle $\theta_c=arcsin((V_0-E)/c\hbar k h_c) $, only helicities of $h \leq h_c$ are coupled in the form of propagating waves.
However, the transmission properties would be modified dramatically in this regime. Besides the cutoff condition inside the barrier, the evanescent waves also exert cutoff conditions for different helicities outside the barrier where a small helicity is transformed into a larger one. Since $sin\phi_{out}^h=sin\theta h/h_0$, for incident wave beyond the critical angle $\theta_c=arcsin(h_0/h)$, the transmitted wave with helicity $h$ become evanescent. This gives a cutoff condition for each helicity component which is larger than the helicity of the incident particle. It is clear that if the incident particle follows the outermost cone of the multicone structure, there will be no cutoff for any helicity outside the barrier.

In Section ~\ref{klein_section} we present  detailed numerical results of Klein birefringent tunneling by evaluating the different transmission coefficients of spin-$3/2$ Dirac-Weyl fermions with a double cone structure. From the results there, the transmission properties of more complex multilayered cone structure can readily be deduced.


\begin{references}
\bibitem{comment}
We note the free Hamiltonian of the Dicke model is absent in Eq.~\eqref{dicke_model}, but  could be 
obtained from the squared momentum and a spin-dependent external trapping potential, 
which will emerge naturally from the dark state BEC system~\cite{review_synthetic_gauge_fields}. Thus a simulation of 
$\mathfrak{su}(2)$ algebra with 
degenerate dark states would give a complete mapping to the single mode Dicke Hamiltonian. These results 
will be reported elsewhere.
\bibitem{comment2}
For compact manifolds, such as the 2-torus associated to the Brillouin zone, all elliptic differential operators are of Fredholm type~\cite{nakahara_book}. Accordingly, they have a finite-dimensional kernel. Since we have shown that  there exists an infinite number of zero modes for integer spin (see Table~\ref{WLL_spectrum}),   this proves that the differential operators in the Hamiltonian are not Fredholm, and thus not elliptic. Under these circumstances, the Atiyah-Singer index theorem does not apply.
\end{references}
\end{document}